\documentclass[11pt,a4paper]{article}

\usepackage{amsmath,amssymb}
\usepackage{epsfig,graphicx}
\usepackage{subfigure}
\usepackage{graphicx}
\usepackage{rotating}
\usepackage{cancel}
\usepackage{bm}
\usepackage{cite}

\newcommand{\higgs}{H_{\rm h}}
\newcommand{\higgst}{\tilde{H}_{ \rm h}}
\renewcommand{\Re}{\mathrm{Re}}

\def\bra{\langle}
\def\ket{\rangle}
\def\beq{\begin{equation}}
\def\eeq{\end{equation}}
\newcommand{\C}[1]{\mathcal{#1}}
\def\ov{\overline}

\begin{document}
\numberwithin{equation}{section}
\title{{\normalsize  IPPP/09/63; DCPT/09/126;  DESY 09-123\hfill\mbox{}\hfill\mbox{}}\\
\vspace{2.5cm} \Large{\textbf{Naturally Light Hidden Photons in\\ LARGE Volume String Compactifications}}}
\author{Mark Goodsell$^{1}$,
\
Joerg Jaeckel$^{2}$,
\
Javier Redondo$^{3}$
\
and Andreas Ringwald$^3$
\\[2ex]
\small{\em $^1$LPTHE, Universit\'e Pierre et Marie Curie - Paris VI, France}\\[1.5ex]
\small{\em $^2$Institute for Particle Physics and Phenomenology, Durham University, Durham DH1 3LE, United Kingdom}\\[1.5ex]
\small{\em $^3$Deutsches Elektronen-Synchrotron DESY, Notkestra\ss e 85, 22607 Hamburg, Germany}\\[1.5ex]
}
\date{}
\maketitle

\begin{abstract}
\noindent Extra ``hidden'' U(1) gauge factors are a generic feature
of string theory that is of particular phenomenological interest.
They can kinetically mix with the Standard
Model photon and are thereby accessible to a wide variety of
astrophysical and cosmological observations and laboratory
experiments. In this paper we investigate the masses and the
kinetic mixing of hidden U(1)s in LARGE volume compactifications of
string theory. We find that in these scenarios the hidden photons
can be naturally light and that their kinetic mixing with the
ordinary electromagnetic photon can be of a size interesting for
near future experiments and observations.
\end{abstract}

\newpage

\section{Introduction}

\begin{figure}[t]
\centerline{\includegraphics[angle=-90,width=16cm]{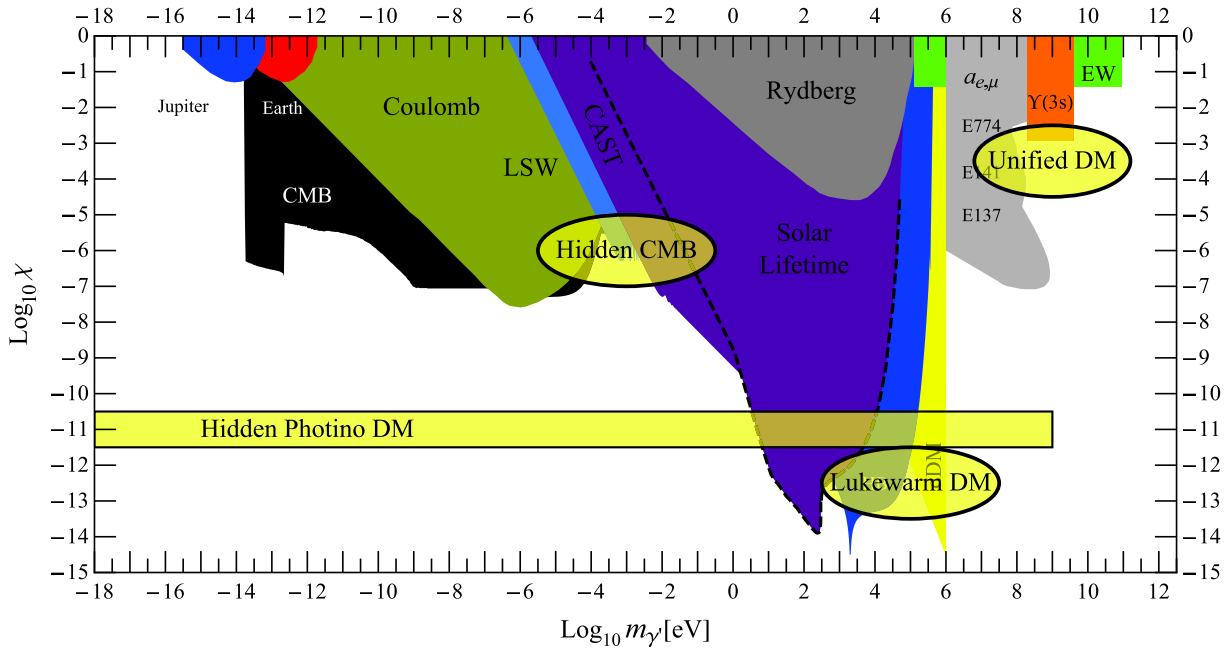}}
\caption{
Current experimental limits on the possible existence of a hidden photon of
mass $m_{\gamma^\prime}$, mixing kinetically with the photon, with a mixing parameter
$\chi$. Strong constraints arise from the non-observation of deviations from
the Coulomb law (yellow)~\cite{Williams:1971ms,Bartlett:1988yy},
from Cosmic Microwave Background (CMB) measurements of the effective number of neutrinos and the blackbody nature of the spectrum (black)~\cite{Jaeckel:2008fi,Mirizzi:2009iz}, from light-shining-through-walls (LSW) experiments (grey)~\cite{Ruoso:1992nx,Cameron:1993mr,Robilliard:2007bq,Ahlers:2007rd,Chou:2007zzc,Ahlers:2007qf,Afanasev:2008jt,Fouche:2008jk,Afanasev:2008fv,Ehret:2009sq}, and from
searches of solar hidden photons with the CAST experiment (purple)~\cite{Andriamonje:2007ew,Redondo:2008aa}.
The white region in parameter space is currently unexplored, but may be accessed by
experiments in the very near future, in particular by improvements in LSW experiments (for proposed experiments probing
this region, see Refs.~\cite{Jaeckel:2007ch,Gninenko:2008pz,Jaeckel:2008sz,Yale,Daresbury,tobar,Caspers:2009cj}). The yellow regions indicate some especially interesting regions as described in the main
text.
}\label{Fig:current_limits}
\end{figure}

Hidden gauge factors, appearing in the low-energy effective field theory,
seem to be a common feature of string theory after
compactification to four space-time dimensions.
The corresponding hidden gauge bosons are very weakly coupled to the
visible sector particles because their mutual interactions are loop-suppressed.
Of particular phenomenological interest are hidden U(1) factors, since hidden photons
may remain very light and therefore may lead to observable effects in astrophysics,
cosmology, or in the laboratory.
Their interactions with the visible photon are encoded in a low-energy
effective Lagrangian of the generic form\footnote{In Eq.~\eqref{LagKM}, we are neglecting explicit higher dimensional operators,
such as $F^4$ and the like, which are suppressed by powers of the string scale. However, we discuss and include below
effective kinetic mixings arising from
inserting vacuum expectation values into suitable higher dimensional operators.}
\begin{equation}
\mathcal{L} \supset -\frac{1}{4g_a^2} F^{(a)}_{\mu \nu} F^{\mu \nu}_{(a)}
- \frac{1}{4g_b^2} F^{(b)}_{\mu \nu} F^{\mu \nu}_{(b)}
+ \frac{\chi_{ab}}{2 g_a g_b} F_{\mu \nu}^{(a)} F^{(b) \mu \nu}
+ \frac{m_{ab}^2}{g_ag_b} A^{(a)}_{\mu} A^{(b)\mu},
\label{LagKM}
\end{equation}
where
$a(b)$ labels the visible (hidden) U(1) gauge field, with
field strength $F^{(a(b))}_{\mu\nu}$ and gauge coupling $g_{a(b)}$.
Clearly, the phenomenological consequences will strongly depend on the
relative strength of kinetic~\cite{Holdom:1985ag},
\begin{equation}
\left( \chi_{ab}\right) = \left( \begin{array}{cc}  0 & \gamma \\  \gamma & 0   \end{array} \right) ,
\end{equation}
and mass mixing~\cite{Okun:1982xi,Babu:1997st,Feldman:2007wj}.
The latter, on account of the apparent masslessness of the photon, should be approximately of the form
\begin{equation}
\left( m^2_{ab}\right) \approx  \left( \begin{array}{cc}  0 & 0 \\  0 & m_{\gamma^\prime}^2 \end{array} \right) .
\label{req_val}
\end{equation}
In fact, phenomenologically very interesting pairs of parameters are (see also Fig.~\ref{Fig:current_limits}):
\begin{enumerate}
\item $(\chi,m_{\gamma^\prime} )\sim (10^{-6}, 0.2\ {\rm meV}$), the region labeled ``Hidden CMB" in
Fig.~\ref{Fig:current_limits},
leading to a natural explanation of the finding of some global cosmological analyses
that present precision cosmological
data on the cosmic microwave background and on the large scale structure of the universe
appear to require some extra radiation energy density from invisible particles apart from the three known neutrino
species~\cite{Jaeckel:2008fi}.
Moreover, these values are accessible to ongoing laboratory experiments exploiting
low energy photons~\cite{Ahlers:2007rd,Jaeckel:2007ch,Ahlers:2007qf,Jaeckel:2008sz,Caspers:2009cj,Ahlers:2008qc}
and allow for interesting technological
applications of hidden photons~\cite{Jaeckel:2009wm}.
\item $(\chi,m_{\gamma^\prime} )\sim (10^{-12}, 0.1\ {\rm MeV}$), the region labeled ``Lukewarm DM" in
Fig.~\ref{Fig:current_limits},
allowing the hidden photon to be a lukewarm dark matter candidate~\cite{Pospelov:2007mp,Redondo:2008ec}.
\item $(\chi,m_{\gamma^\prime} )\sim (10^{-4},  {\rm GeV}$), the region labeled ``Unified DM" in
Fig.~\ref{Fig:current_limits}. For these values, the hidden photon
plays an important role in models where the dark matter resides in the
hidden sector~\cite{ArkaniHamed:2008qn}. These models aim at a unified description of unexpected observations
in astroparticle physics, notably the positron excess observed by
the satellite experiment PAMELA \cite{Adriani:2008zr} and the annual modulation signal seen by the direct dark matter search experiment DAMA~\cite{Bernabei:2008yi}.
The massive hidden U(1) can then mediate ``Dark Forces''.
These values are also accessible to accelerator searches~\cite{ArkaniHamed:2008qp,Baumgart:2009tn,Essig:2009nc,Bjorken:2009mm}
and have been motivated in various supersymmetric scenarios~\cite{Chun:2008by,Cheung:2009qd,Morrissey:2009ur,Baumgart:2009tn,Cui:2009xq}.
See also Ref.~\cite{Suematsu:2006wh,Katz:2009qq,Batell:2009yf,Batell:2009di,Dedes:2009bk,Ruderman:2009tj}.
\item $(\chi,m_{\gamma^\prime} )\sim (10^{-11}, \lesssim  100\,{\rm GeV}$), the region labeled ``Hidden
Photino DM" in Fig.~\ref{Fig:current_limits}. For these values the supersymmetric partner
of the hidden photon, the hidden photino, is a promising dark matter candidate, if its mass is
in the 10 to 150 GeV range~\cite{Ibarra:2008kn}.
\item $(\chi,m_{\gamma^\prime} )\sim (10^{-23}, 0$), in which case the hidden photino, with mass in the TeV range, may be a candidate of decaying dark matter, giving rise to
the above mentioned excesses observed in galactic cosmic
ray positrons and electrons~\cite{Shirai:2009kh,Ibarra:2009bm}. A phenomenologically quite similar scenario is obtained when
the hidden photon itself is massive, in the TeV range, and decays into Standard Model
particles through kinetic mixing with a U(1)$_{B-L}$ gauge boson, with
$\chi_{\gamma^\prime,B-L}\sim 0.01$ and $m_{B-L}\sim 10^{15}\ {\rm GeV}$~\cite{Chen:2008md,Chen:2008qs}.
\item $(\chi,m_{\gamma^\prime} )\sim (10^{-3}, {\rm TeV}$). This  is phenomenologically interesting because the hidden photon may
be probed at colliders~\cite{Feldman:2007wj,Kumar:2006gm,Chang:2006fp,Coriano:2007xg,Kors:2004ri}.
\end{enumerate}

We see that a wide range of values for the kinetic mixing parameter and the mass of the
hidden photon leads in fact to very interesting physics.
Therefore, it seems timely to investigate which values are naturally obtained in
realistic string compactifications. Intriguingly, in compactifications of
type II strings, the possible values for kinetic mixing are indeed widespread, reflecting
the diverse possibilities of sizes and fluxes in the compactified dimensions~\cite{Abel:2003ue,Abel:2006qt,Abel:2008ai}. We thus concentrate our
investigations in this paper to predictions from type II theories\footnote{First calculations of kinetic mixing
in toroidal compactifications of type II
and type I strings have appeared in
\cite{Lust:2003ky,Abel:2003ue,Berg:2004ek},
respectively. For analyses of kinetic mixing in compactifications of the heterotic string, see \cite{Dienes:1996zr,Lukas:1999nh,Blumenhagen:2005ga}.}, in particular to the
ones with possibly large bulk volumes~\cite{Balasubramanian:2005zx,Conlon:2005ki}, which seem
to be very successful phenomenologically~\cite{Conlon:2008wa}.

The paper is set up as follows. In the following Sect.~\ref{hyperweak} we will briefly review the setup of LARGE volume scenarios and
hyperweak gauge interactions. In Sect.~\ref{mixing} we will determine the expected kinetic mixing in these scenarios.
Since we are interested in massive hidden U(1) gauge bosons we will discuss mechanisms to generate masses in Sects.~\ref{Stueck} and \ref{Higgs}.
Finally, we collect and summarize our results in Sect.~\ref{conclusions}.

\section{LARGE volumes and hyperweak interactions}\label{hyperweak}

\begin{center}\begin{figure}
\begin{center}
\subfigure[]{
\includegraphics[angle=-90,width=7.7cm]{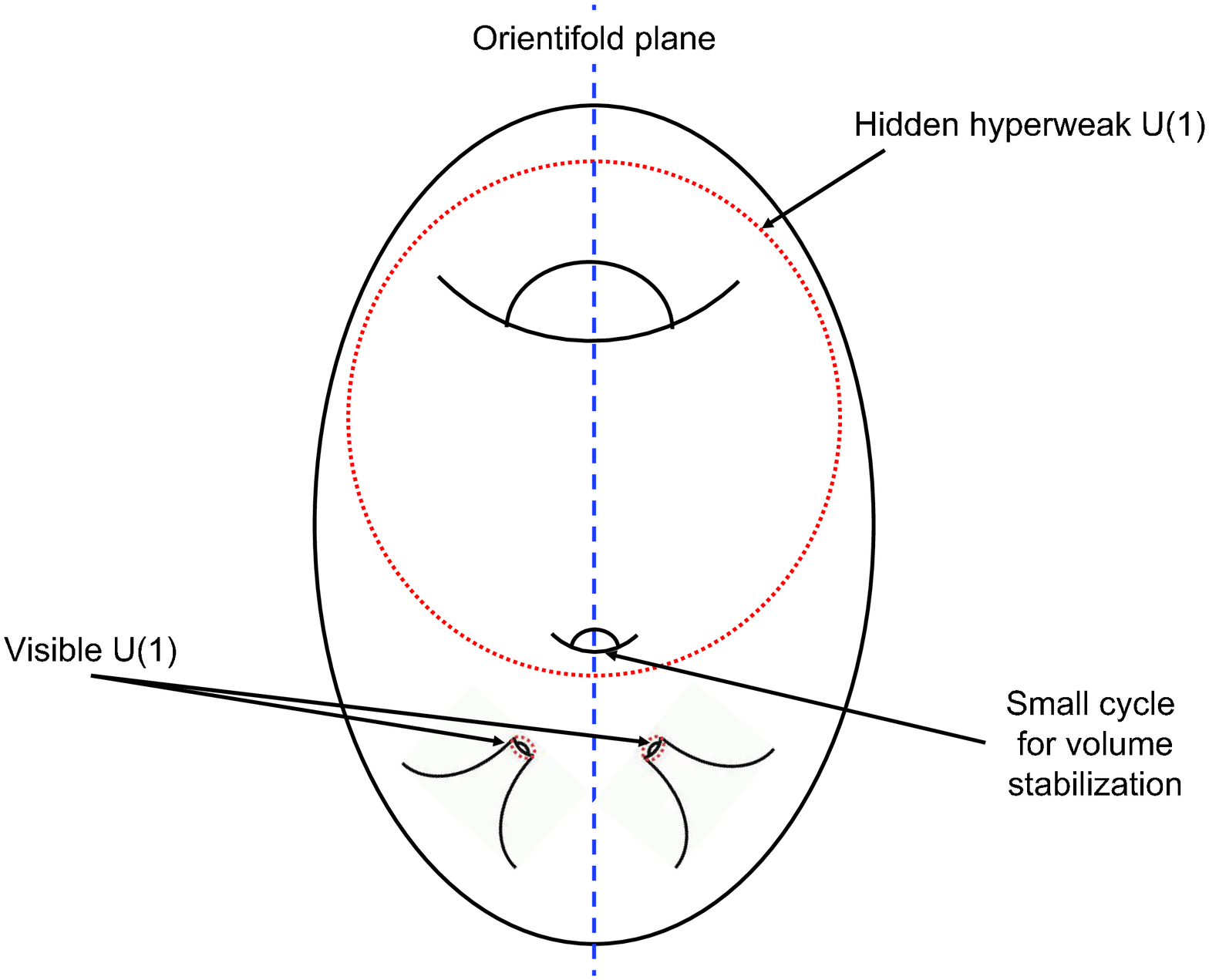}\label{scenario1}}
\subfigure[]{
\includegraphics[angle=-90,width=7.7cm]{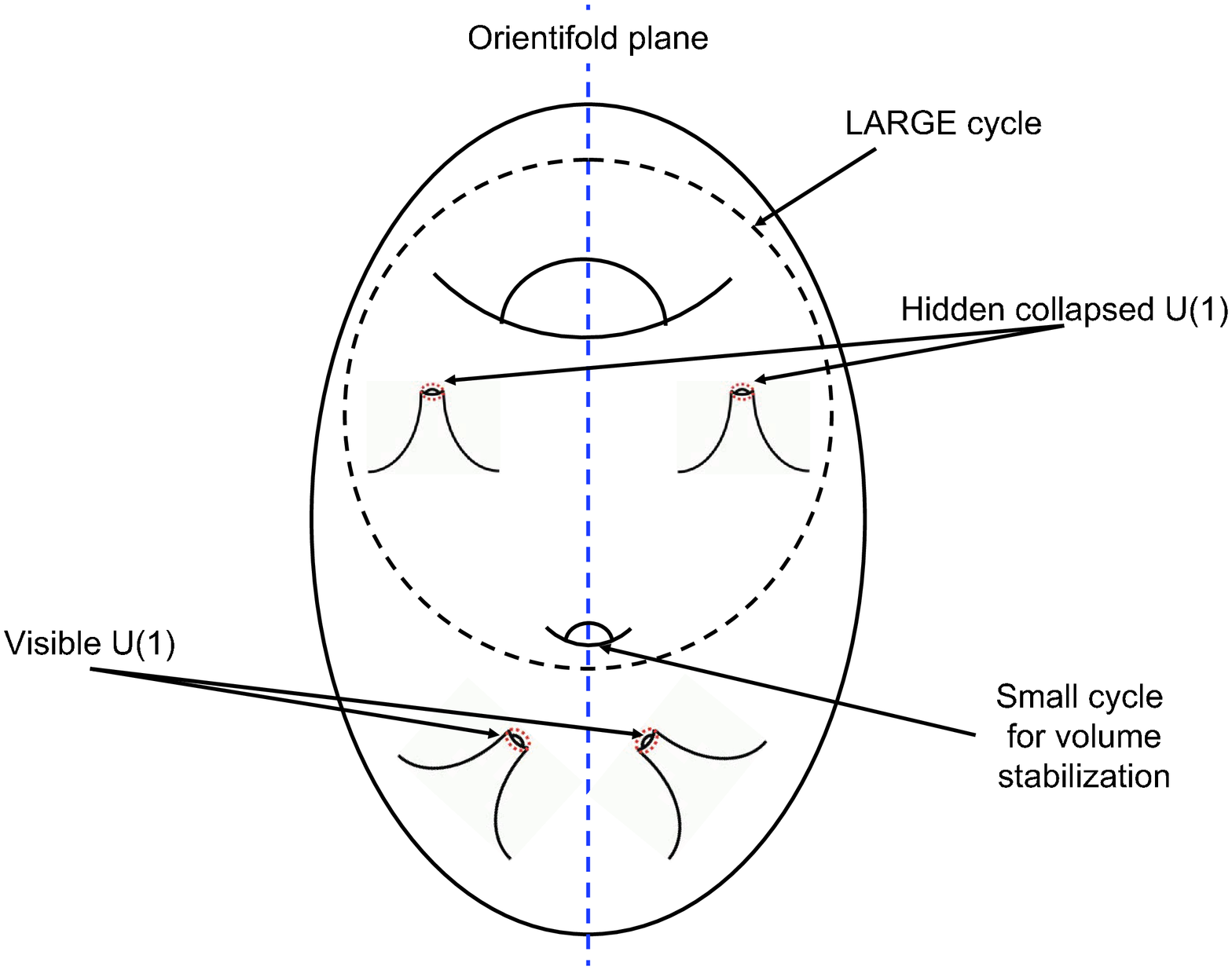}\label{scenario2}}
\subfigure[]{
\includegraphics[angle=-90,width=7.7cm]{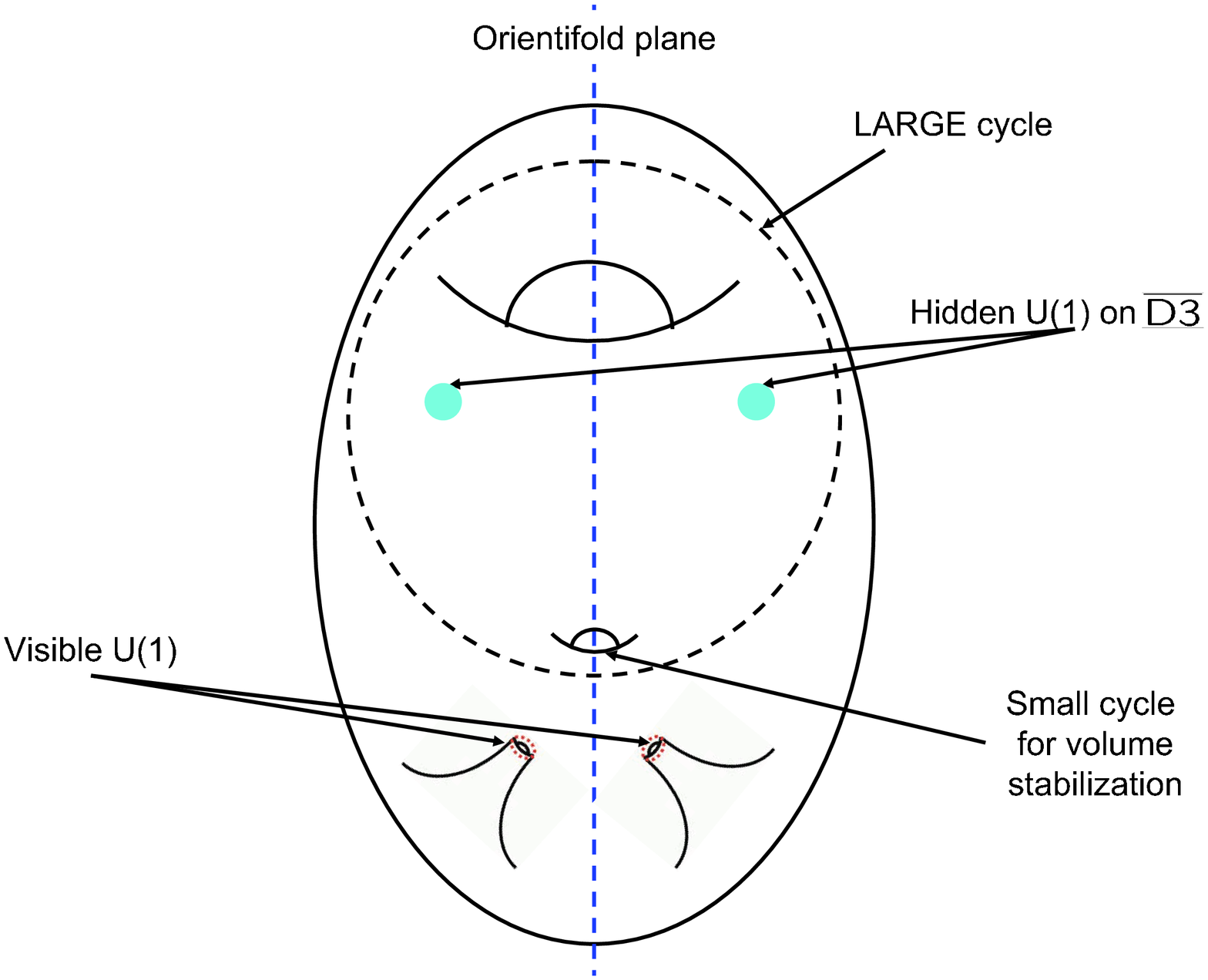}\label{scenario3}}
\end{center}
\caption{Hidden U(1)s in LARGE volume scenarios realized in type IIB orientifold flux compactifications.
A common feature of the geometry of the extra dimensions in these scenarios is that they have a minimum of four cycles:
a large one to control the overall volume, a small one to allow for non-perturbative effects stabilising the large volume,
and two small cycles, exchanged by the orientifold, wrapped by the
visible branes~\cite{Conlon:2008wa}.
This leaves various possibilities for
hidden U(1)s. In \ref{scenario1} the hidden U(1) gauge group is located on a LARGE cycle extending through
the full LARGE volume. In \ref{scenario2} the hidden U(1) is located on a collapsed cycle. The black dashed line indicates the existence of a LARGE cycle.
We do not necessarily have a brane wrapping around this cycle. Finally, in the last scenario \ref{scenario3} the hidden U(1) sits
on an anti D3-brane which is often exploited for uplifting to a de Sitter vacuum~\cite{Kachru:2003aw}.
}\label{scenarios}
\end{figure}\end{center}

LARGE volume theories are based upon type IIB strings with D3 and D7-branes\footnote{In Appendix \ref{APPENDIX:EFFECTIVE} we review the
four-dimensional supergravity corresponding to these theories, its Kaluza-Klein reduction and the identification of moduli.}, which give rise to ``brane worlds". In these theories, the visible sector lives on a stack
of D-branes which are extended along the 3+1 non-compact dimensions and wrap small collapsed
cycles in the compactification manifold (see Fig.~\ref{scenarios}),
while gravity propagates in the bulk, leading to a possibly smaller string scale at the
expense of a larger compactification volume.
In fact, the relation between the (reduced) Planck
scale $M_P=2.4\times 10^{18}$~GeV, the string scale $M_s$, the string coupling $g_s$ and the
total volume $\mathcal{V}\equiv V_6 M_s^6$ of the bulk\footnote{In our conventions, the string scale is $M_s=1/l_s  = 1/(2\pi \sqrt{\alpha^\prime})$,
in terms of the string tension $\alpha^\prime$.} is given by
\begin{equation}
M_P^2 = \frac{4\pi}{g_s^2} \mathcal{V} M_s^2.
\label{MPMs}
\end{equation}
This equation can be read off from the IIB supergravity action in 10 dimensions, Eq.~\eqref{EFFECTIVE:10DIIB} in Appendix \ref{APPENDIX:EFFECTIVE}.
It gives rise to a string scale of order the GUT scale,
$M_s\sim 10^{16}$~GeV, for $\mathcal V \sim 50$, of order
the intermediate scale, $M_s\sim 10^{10}$~GeV, for $\mathcal V\sim 5\times 10^{13}$, and of order the TeV scale, for
$\mathcal V\sim 5\times 10^{27}$.
We will consider the full range of values of the string scale.
We will not address issues such as the lightness of moduli for low string scales.
Nevertheless, it is worth pointing out that there have been recent speculations claiming that even for TeV strings it is possible to avoid the moduli problems~\cite{TeVmoduli}.

%
\begin{center}\begin{figure}
\centerline{\includegraphics[angle=-90,width=8cm]{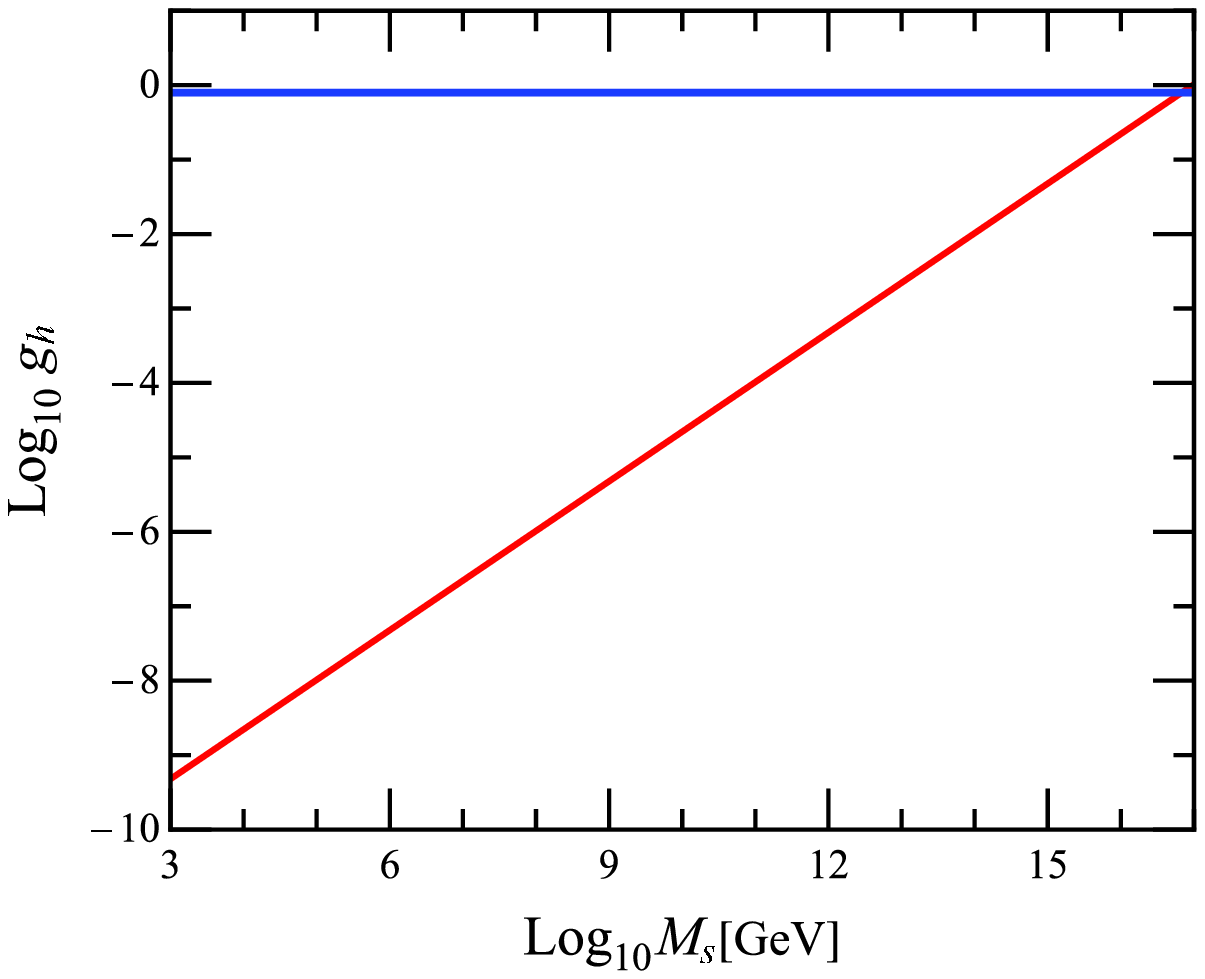}
}
\caption{Gauge coupling $g_{(q)}$ of a hidden U(1)
as a function of the string scale $M_s$, for different
dimensions of the wrapping cycle of the brane hosting the hidden U(1):
$q=0$ (blue) corresponding to an (anti) D3 brane (cf. Fig.~\ref{scenario3}) or a collapsed D7 brane (cf. Fig.~\ref{scenario2}), and
$q=4$ (red) for a D7 brane (cf. Fig.~\ref{scenario1}).
The string coupling has been set to $g_s=0.1$, such that $g_{(0)}$ corresponds to the
hypercharge gauge coupling at the string scale, $\alpha_Y(M_s)\equiv g_Y^2(M_s)/(4\pi )\approx 1/20$.
}\label{Fig:couplingsvsms}
\end{figure}\end{center}

Hidden U(1)s arise in these scenarios from space-time filling D7-branes wrapping cycles
in the extra dimensions which are not intersecting the visible sector branes
(cf. Fig.~\ref{scenarios}(a) and (b)).
Another possibility
giving rise to hidden sector U(1)s are anti D3-branes (cf.  Fig.~\ref{scenarios}(c)).
More precisely, we can place our hidden U(1) on a D7 brane
wrapping a LARGE cycle as in Fig.~\ref{scenario1}. As we will see below, this corresponds to a situation where
the hidden U(1) becomes hyperweakly coupled. We can also put an extra U(1) on collapsed cycles as in Fig.~\ref{scenario2}
or on anti D3 branes as in Fig.~\ref{scenario3}. In both these cases the gauge coupling will be ${\mathcal{O}}(1)$.

In fact, the gauge couplings in these scenarios can be read off from the Dirac-Born-Infeld (DBI) action
(cf. the first term in Eq.~(\ref{EFFECTIVE:DBI})).  The gauge coupling of a U(1) gauge boson
on a D$(3+q)$-brane, wrapping a $q$-cycle of volume $V_q$ in the extra dimensions,
turns out to be related to the volume $\mathcal{V}_{q}=V_{q}/l_{s}^q$ in string units by
\begin{equation}
g_{(q)}^2 = \frac{2\pi g_s}{|Z|} \simeq
\frac{2\pi g_s}{V_{q}M_s^q}\equiv\frac{2\pi g_s}{\mathcal{V}_{q}} \,.
\label{alpha_p}
\end{equation}
Here $|Z|$ is the absolute value of the central charge of the
branes, equal to the right hand side in the large volume limit.
Therefore, in scenarios where the volume $\mathcal{V}_q$ is large, the corresponding
gauge coupling is small, giving rise to hyperweak
interactions~\cite{Burgess:2008ri}.
In fact, from Eqs.~\eqref{MPMs} and \eqref{alpha_p} we estimate
\begin{equation}
g_{(q)}^2 \simeq
\frac{2\pi g_s}{(\mathcal{V})^{q/6}}
= 2\pi g_s \left( \frac{4\pi}{g_s^2} \frac{M_s^2}{M_P^2}\right)^{q/6}\,,
\label{gsquaredq}
\end{equation}
which can be tiny for $0< q\leq 6$ and string scales much below the Planck scale, corresponding
to large bulk volumes.
This can be clearly seen by looking at the red curve in Fig.~\ref{Fig:couplingsvsms},
where we show the estimate for the gauge coupling of a U(1) on a D7 brane.
In contrast the gauge coupling on an (anti) D3 brane or a D7 brane on a collapsed cycle is unaffected by the volume suppression and therefore
comparatively large, i.e. ${\mathcal{O}}(1)$, as can be seen from the blue line in Fig.~\ref{Fig:couplingsvsms}.

\section{Kinetic mixing}\label{mixing}
Prior to the breaking of supersymmetry, in the 4D effective theory the kinetic mixing appears as a holomorphic quantity in the gauge kinetic part of
the supergravity Lagrangian,
\begin{equation}
\mathcal{L} \supset \int d^2 \theta  \left\{
\frac{1}{4 (g_a^h)^2} W_a W_a + \frac{1}{4(g_b^h)^2} W_b W_b
- \frac{1}{2}\chi_{ab}^h W_a W_b  \right\},
\end{equation}
where $W_a, W_b$ are the field strength superfields for the two U(1) gauge fields and $\chi_{ab}^h,g_a^h, g_b^h$ are the holomorphic kinetic mixing parameter and gauge couplings that must run only at one loop. The well known expression for the holomorphic gauge running is
\beq
\frac{1}{(g_a^h)^2(p)} = \frac{1}{(g_a^h)^2(\Lambda)} - \sum_r \frac{n_r Q_a^2 (r)}{8\pi^2} \log p/\Lambda,
\eeq
Here, $Q_a (r) $ denotes the charge under group $a$ carried by $n_r$ fields.
The physical gauge couplings are given in terms of the holomorphic quantities by the Kaplunovsky-Louis formula \cite{Kaplunovsky:1994fg,Kaplunovsky:1995jw} (given
here specialised to U(1) gauge groups):
\beq
g_a^{-2} = \Re\bigg[(g_a^h)^{-2}\bigg] - \sum_r \frac{ Q_a^2 (r)}{8\pi^2} \log \det Z^{(r)} - \sum_r \frac{n_r Q_a^2 (r)}{16\pi^2} \kappa^2 K,
\eeq
where $Z^{(r)}$ is the renormalised kinetic energy matrix of fields having charge $Q_a (r)$ (i.e. the renormalised K\"ahler metric $K_{\ov{\alpha} \beta}$), $K$ is the full K\"ahler potential and $\kappa^2=1/M_P^2$.

We should expect that a similar formula should be obeyed for the kinetic mixing, too. Indeed, it can be shown that the relevant expression, exact to all orders in perturbation theory, is
\beq
\frac{\chi_{ab}}{g_a g_b} = \Re(\chi_{ab}^h) + \frac{1}{8\pi^2} \mathrm{tr}\bigg( Q_a Q_b \log Z\bigg) +\frac{1}{16\pi^2} \sum_r n_r Q_a Q_b (r)\kappa^2 K,
\label{KLTypeFormula}\eeq
where $\chi_{ab}$ is now the parameter in the canonical  Lagrangian density
\beq
\C{L}_{\mathrm{canonical}} \supset \int d^2 \theta \left\{
\frac{1}{4} W_a W_a + \frac{1}{4} W_b W_b
- \frac{1}{2}\chi_{ab} W_a W_b  \right\}.
\eeq
This is easiest to understand using the techniques of \cite{ArkaniHamed:1997mj}; it is a consequence of rescaling the vector superfields from the holomorphic basis (where the gauge kinetic terms are as above) to the canonical basis, $V_a \rightarrow g_a V_a$. In fact, the analysis used there follows through exactly for the kinetic mixing (with the exception of the supergravity contribution proportional to the full K\"ahler potential).
Apart from the applications in this paper, the above formula can be used to derive the running of couplings for several U(1) factors to specified orders in perturbation theory, such as in \cite{delAguila:1988jz,Babu:1996vt}.

It is worth also noting that the same formalism also describes ``magnetic mixing''~\cite{Brummer:2009cs,Bruemmer:2009ky}. Defining $\mathrm{Im} \left( \frac{1}{(g_a^h)^2} \right) = \frac{\theta_a}{8\pi^2} $ and similarly for $\theta_b$, there are terms in the Lagrangian density
\beq
\C{L} \supset \frac{\theta_a}{16\pi^2} F_{\mu \nu}^a\tilde{F} ^{a,\,\mu \nu} + \frac{\theta_b}{16\pi^2} F_{\mu \nu}^b \tilde{F} ^{b\,\mu \nu} - \frac{\theta_{ab}}{8\pi^2} F_{\mu \nu}^a \tilde{F} ^{b\,\mu \nu},
\eeq
where a tilde denotes the dual field strength, and $\theta_{ab}$ is the ``magnetic-mixing'' angle,
\beq
\mathrm{Im} ( \chi_{ab} ) = \frac{\theta_{ab}}{8\pi^2}.
\eeq
We then find a ``physical'' magnetic mixing in analogy to the above; the resulting canonical Lagrangian density is \cite{Benakli:2009mk}
\beq
\C{L}_{\mathrm{canonical}} \supset  \frac{g_a^2 \theta_a}{16\pi^2} F_{\mu \nu}^a\tilde{F}^{a \mu \nu} + \frac{(g_b)^2 \theta_b}{16\pi^2} F_{\mu \nu}^b \tilde{F} ^{b\,\mu \nu} - \frac{g_a g_b \theta_{ab}}{8\pi^2} F_{\mu \nu}^a \tilde{F}^{b\,\mu \nu}.
\eeq
Since $\chi^h_{ab}$ run only at one loop, the same is true of $\theta_a, \theta_b, \theta_{ab}$, with the ``physical'' quantities obtaining corrections only through the modification of the gauge coupling.

So far we have considered the form of the gauge coupling in 4D supergravity.
Further information can be obtained by dimensional reduction from the original 10 dimensions. In particular, moduli fields will now appear.
Again, we will specialize to the case of a IIB setup.

As reviewed in Appendix \ref{APPENDIX:EFFECTIVE}, in LARGE volume type IIB models, the gauge groups are supported on D7 branes, which wrap four-cycles $\tau_i$ corresponding
to the $h^{2}$ K\"ahler moduli $T_i$,
but may in addition support magnetic fluxes wrapping two-cycles $t_i$.

$\chi_{ab}^h$ depends upon the moduli, both closed and open. The closed string K\"ahler moduli $T_\alpha$ transform under Peccei-Quinn symmetries
(which lead to discrete shifts of their imaginary parts). Therefore, they can only enter as exponentials.
However, they also depend upon the inverse string coupling \cite{Grimm:2004uq,Grimm:2005fa} (see equation
(\ref{EFFECTIVE:Kaehlerdef}) and the text below it)
and consequently an exponential
dependence would be non-perturbative. Thus they \emph{cannot} enter at 1-loop. Accordingly, $\chi_{ab}$ is given by
\begin{equation}
\chi_{ab}^h = \chi_{ab}^{\mathrm{1-loop}} (z^k, y_i) + \chi_{ab}^{\mathrm{non-perturbative}} (z^k, e^{-T_j}, y_i) ,
\end{equation}
where $z^k$ are the complex structure moduli, and $y_i$ are the open string moduli. The above is in analogy to the structure of gauge kinetic functions,
see for example \cite{Akerblom:2007uc}. Generically, the $z^k$  enter the holomorphic kinetic mixing in polynomial or exponential form (for example in the explicit example of toroidal models they enter via powers of exponentials~\cite{Abel:2008ai}) and will typically be numbers of order one, although certain may be exponentially small at the end of a warped throat~\cite{Giddings:2001yu}. We thus conclude that generically
\begin{equation}
\chi_{ab}^h \sim \frac{1}{16\pi^2}.
\end{equation}
Using the above expression (\ref{KLTypeFormula}) to compute the canonical kinetic mixing and the fact that, for a hidden U(1) separated by
distances greater than the string scale from the hypercharge there are no light states connecting them (i.e. charged under  both)
and thus no contributions from the K\"ahler potential,
we expect a kinetic mixing of order
\begin{equation}
\chi_{ab} \sim \frac{g_a g_b}{16\pi^2} \ .
\label{SUSYKMEstimate}
\end{equation}

\begin{center}\begin{figure}
\centerline{
\includegraphics[angle=-90,width=8cm]{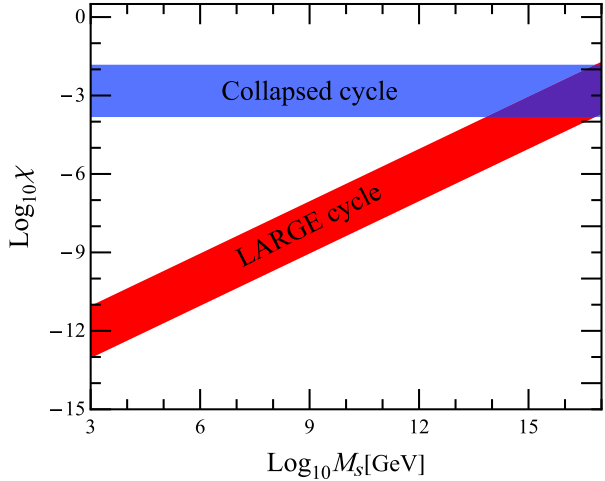}}
\caption{
Kinetic mixing between the visible electromagnetic U(1) and a U(1) sitting on a collapsed cycle (blue) or
a hyperweak U(1) on a LARGE cycle (red), as a function of the string scale.}
\label{Fig:chivsms}
\end{figure}\end{center}

This is one of our main results. In Fig.~\ref{Fig:chivsms}, we display the predictions of
kinetic mixing for the two first scenarios of hidden U(1)s shown in Fig.~\ref{scenarios}, as a
function of the string scale, which we vary from the TeV scale up to the GUT scale.
For the case of a hyperweak U(1), realized by the geometric setup in Fig.~\ref{scenarios} (a),
quite interesting values for kinetic mixing are predicted,
\begin{equation}
\chi_{ab}  \sim
\frac{2\pi g_s}{16 \pi^2} \left( \frac{4\pi}{g_s^2} \frac{M_s^2}{M_P^2}\right)^{1/3}\,,
\end{equation}
as is apparent from the red band
in Fig.~\ref{Fig:chivsms} and a comparison with the present phenomenological limits in
Fig.~\ref{Fig:current_limits}. Even larger values are predicted if the hidden U(1) sits
on a collapsed cycle, as in Fig.~\ref{scenarios} (b), because the hidden gauge coupling is in this
case just of order one, irrespective of the string scale.
The corresponding estimate is the blue band in Fig.~\ref{Fig:chivsms}.

So far, we have discussed the generic expectation for kinetic mixing between the visible
U(1) and a hidden U(1) in LARGE volume string compactifications.
Much smaller values could result in special cases where the one-loop contribution is cancelled or vanishes.
They arise from a non-perturbative contribution which would be suppressed by a factor of $e^{-aT}$ for some $a, T$.
In the following subsection \ref{special} we will therefore briefly check when this could happen.

Another suppression of the mixing could result from effects from the K\"ahler potential corrections in~\eqref{KLTypeFormula}.
Those become effective, however, only in the case when the hyperweak brane intersects the Standard Model brane and there are states
charged under both hypercharge and the hyperweak group.

\subsection{Special cases}\label{special}

We argued previously that the generic value for the holomorphic kinetic mixing parameter is $\mathcal{O}\left(\frac{1}{16\pi^2}\right)$.
However, there are certain cases where it may vanish, and we must carefully examine the details of the compactification.

Firstly there exists the simplest case, that we have a toroidal model with parallel branes without magnetic fluxes (this forced the calculations of \cite{Abel:2003ue,Abel:2006qt} to specialise to the mixing between branes and antibranes). In this case it is the fact that the messenger fields stretching between the branes fall into $N=4$ multiplets that effects the cancellation (recall that the calculation is similar to that of gauge threshold corrections, and the cancellation thus follows from the vanishing beta function for $N=4$ supersymmetry). However, this is somewhat misleading, since on a general Calabi-Yau manifold (rather than an orbifold of the torus) there can only be two supersymmetries, and we can therefore assume that even in the case of parallel branes in such cases we will have non-vanishing masses and mixings.

The second case is cancellations due to orientifold images. If we are calculating the mixing between two branes $a$ and $b$ and their images in the upstairs geometry $a'$ and $b'$, then for USp-type projections the mixing will be $\chi_{ab} - \chi_{ab'} - \chi_{a'b} + \chi_{a'b'} = 2 (\chi_{ab} - \chi_{ab'}) = 2(\chi_{ab} - \chi_{a'b})$ (for SO-type projections the signs will all be positive and we can thus expect a mixing unless the branes lie on the orientifold planes). Clearly if either brane $a$ or $b$ lies on the orientifold plane and its gauge bundle is invariant then this will vanish. Other than this case and the special case of toroidal models, however, we generically expect $\chi_{ab} \ne \chi_{ab'}$. In fact, as discussed in \cite{Abel:2008ai},
the technique of displacing away from orientifold planes merely cancels the mass for the U(1), whilst allowing
for kinetic mixing.

The third, final and most important case is that the massless U(1), i.e. the visible U(1), crucially resides on a single stack of
branes broken by giving a vacuum expectation value (vev) to one type of magnetic flux, such as that considered in local IIB and
F-theory GUTs~\cite{Blumenhagen:2008zz,Beasley:2008dc,Donagi:2008kj,Blumenhagen:2009gk}. For example,
with a stack of five D7-branes wrapping a divisor $D$ and making an SU(5) group, we can turn on a flux vev for one element
of $H^{1,1} (D)$ (trivial on the overall Calabi-Yau to ensure the U(1) is massless) proportional
to $\mathrm{diag}(2,2,2,-3,-3)$ in the gauge indices, which breaks the model to SU(3)$\times$SU(2)$\times$U(1).
We can view the U(1) as two U(1)s with field strengths $F^2_{\mu \nu}\frac{1}{6} \mathrm{diag} (2,2,2,0,0)$ and $F^3_{\mu \nu} \frac{1}{6} \mathrm{diag} (0,0,0,-3,-3)$. An additional non-trivial flux can be turned on to make the
trace-U(1) massive.

In these models, we can analyse the closed string fields that mediate the mixing and determine whether it will vanish. The relevant couplings arise from the dimensional reduction of the Chern-Simons action in Eqs. (\ref{EFFECTIVE:DBI}),(\ref{EFFECTIVE:SDBI})\footnote{We have neglected the curvature terms for simplicity. The reader may confirm that, since they do not carry gauge indices, they will not affect the discussion below.} and are given by
\begin{align}
\label{interaction}
\frac{  M_s^2}{ g_s \pi} \bigg[&\int_{{\rm D}7_{\rm vis}} F_{\rm vis} \wedge \star_4 B_2 \wedge\frac{1}{2} \bigg(J \wedge J - c_{1}(\C{L_{\rm vis}})^2\bigg)\nonumber \\
+&\int_{{\rm D}7_{\rm vis}} F_{\rm vis} \wedge  E_2^{(2)} \wedge  c_{1}(\C{L_{\rm vis}})^2 \nonumber \\
+&\int_{{\rm D}7_{\rm vis}} F_{\rm vis} \wedge  E_2^{(4)} \wedge Z_{2} \wedge c_{1} (\C{L_{\rm vis}}) \nonumber \\
+&\int_{{\rm D}7_{\rm vis}} F_{\rm vis} \wedge  E_2^{(6)} \wedge Z_{4}\bigg]\nonumber\\
+&[{\rm vis}\leftrightarrow {\rm hidden}] \ ,
\end{align}
where $\star_4, \star$ are the four and, for future use, six-dimensional Hodge star operators respectively, $J$ is the K\"ahler form on the
compact dimensions, we have defined $E_2^{(i)}$ to be the two-form 4-d ($x$) component of the $C_i$  R-R $i$-forms, and $Z_j$ are
the $j$ form 6-d ($y$) components; $C_i = E_2^{(i)}(x) \wedge Z_{i-2} (y)$. $c_1({\mathcal{L}})$ is the first Chern
class of the line bundle ${\mathcal{L}}$. It is equal to the curvature two-form divided by $2\pi$ and this is equal to the full
higher dimensional fields strength. On the visible brane the GUT group is broken by $ c_1 (\C{L}) \ne 0$ and we find $\int c_1^2 (\C{L}) \ne 0$.

We are now interested in the cross terms between the visible and the hidden parts of Eq.~\eqref{interaction}.
Since the GUT group is broken only by the fluxes, any contribution that is not dependent on the internal fluxes will
cancel due to the tracelessness of the hypercharge. This removes the six-form and the $J \wedge J$ piece coupling to the antisymmetric tensor $B_2$.
Then we can see that the remaining $B_2$ piece exactly cancels with the $C_2$ contribution since they are scalars on the Calabi-Yau. This leaves only two potential contributions: where the $B_2$ field couples to the flux on the visible brane but $\frac{1}{2} J \wedge J$ on the hidden brane (i.e. only applicable for hidden
photons with non-zero Chan-Paton traces on at least one stack of branes) and that coming from the axion $E_2^{(4)}$.

The D7 brane wraps a divisor $D$ in the compact space, and once we have introduced a GUT-breaking flux, the integral $\int_{D} c_1^2 (\C{L}) \ne 0$
(this contributes to the D3-brane tadpole), every component of the $B_2$ field that is non-vanishing at the two-cycle dual to $c_1 (\C{L})$ couples
it to branes with non-vanishing Chan-Paton trace (note that the two-cycle must have zero volume if the manifold is K\"ahler).
Thus we expect our hyperweak gauge group to have a kinetic mixing with the hypercharge provided it does not have a similar GUT-breaking flux,
since it does not lie on an orientifold plane (otherwise it would not carry a U(1)) .

To establish whether the final contribution will vanish, i.e. whether the hypercharge will mix with hidden U(1)s at singularities similar to the visible one, we must understand the propagation of the R-R forms on the Calabi-Yau space. The calculation of kinetic mixing can be interpreted as the integrating out of the form by its equation of motion; we wish to solve
\beq
(k^2  + \star d \star d) Z_2 = A_1^\mu k_\mu \star\bigg( [D_1] \wedge c_1 (\C{L}_1)\bigg) +  A_2^\mu k_\mu \star\bigg( [D_2] \wedge c_1 (\C{L}_2)\bigg)
\eeq
and extract the piece proportional to $k^2$ (since the zero mode gives us the masses). In fact, the above should reduce to finding the Green function of the Laplacian on the compact space once we take into account gauge fixing of the forms under $C_p \rightarrow C_p + d\eta_{p-1}$. We then need to examine the eigenforms of the Laplacian $(d\star d \star + \star d \star d) Z_2 = \lambda Z_2$. A given eigenform will then mediate mixing if $\int_{\alpha_1 \subset D_1} Z_2 \ne 0, \int_{\alpha_2 \subset D_2}Z_2 \ne 0$, where $\alpha_1, \alpha_2$ are the two-cycles on the divisors $D_1, D_2$ dual to $c_1(\C{L})$. If $\alpha_1 = \partial a$ is trivial, then $\int_{\alpha_1 \subset D_1} Z_2 = \int_a d Z_2$, so we expect non-zero modes to contribute. However, since $\alpha_{1,2}$ have zero volume on a K\"ahler manifold, this requires $Z_2$ to be singular. This implies that generically this contribution will be zero unless $\partial a = \alpha_1 + \alpha_2$ (similar to the setup of \cite{Grimm:2008ed}) i.e. the three-cycle $a$ should extend into the bulk and connect the two hidden sectors. This is similar to gauge threshold corrections \cite{Conlon:2009qa}.

In summary, we generically expect a non-vanishing kinetic mixing between the visible U(1) and a hyperweak one (scenario shown in Fig.~\ref{scenario1}).
In the mixing between the visible U(1) and a hidden U(1) on a collapsed cycle (cf. Fig.~\ref{scenario2}) cancellations may occur but there are also setups of this
type where we expect non-vanishing kinetic mixing.

\subsection{\cancel{SUSY} contributions}\label{susybreak}

\begin{center}\begin{figure}
\centerline{
\includegraphics[angle=-90,width=8cm]{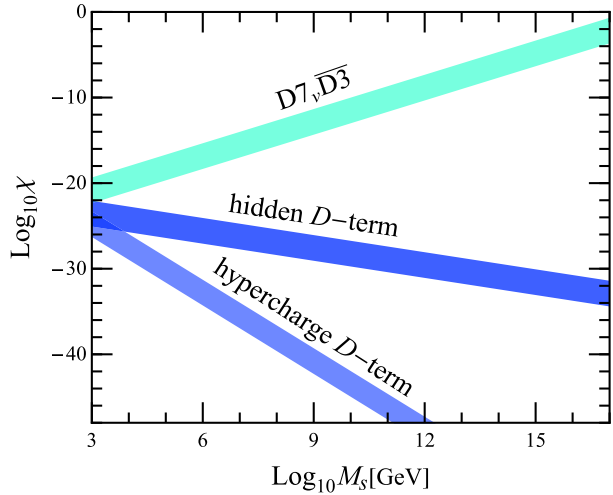}
}
\caption{
Contributions from SUSY breaking to the kinetic mixing between the visible electromagnetic U(1) and a U(1) on a hidden $\ov{\rm D3}$ brane (turquoise) or a hidden collapsed brane through non-vanishing $D$-terms in the electroweak sector (light blue) or in the hidden sector (dark blue).
}
\label{Fig:chivsms_susybroken}
\end{figure}\end{center}

In addition to supersymmetric kinetic mixing contributions, there will also be those coming from supersymmetry breaking effects. The first kind, extensively discussed in \cite{Abel:2008ai}, is that from mixing between branes and anti-branes. In LARGE volume models where the uplift to a de Sitter vacuum is
provided by $\ov{\rm D3}$-branes (cf. Ref.~\cite{Kachru:2003aw} and Fig.~\ref{scenarios} (c)),
the predicted hidden U(1)s
will generically mix with U(1)s with a term that is volume suppressed but may not obtain a St\"uckelberg mass
(cf. Sect.~\ref{Stueck}) since the $B_2$ and $C_2$ zero modes are projected out. From the analysis above, we can see that these will mix with the hypercharge even in a GUT model, since the $B_2$ and $C_2$ contributions now have the same sign and couple to $c_1 (\C{L})^2$ on the visible D7 brane. The mixing thus generically has magnitude \cite{Abel:2003ue}
\beq
\chi_{{\rm D}7_v\ov{\rm D3}} \sim \frac{2\pi g_s}{16\pi^2} \frac{1}{\C{V}^{2/3}} ,
\eeq
giving $\chi \sim 1\times 10^{-21}$ for TeV strings and $3\times 10^{-4}$ for $M_s\sim 10^{16}$ GeV.
This is shown as the turquoise band in Fig.~\ref{Fig:chivsms_susybroken}.

It should be emphasized here that, for TeV strings, this prediction fits very well to the
region phenomenologically favored by a possible explanation of excesses in galactic cosmic ray data
through decaying hidden photino dark matter
(item 5 in the Introduction).
Since uplifting to a de Sitter space is a phenomenological necessity, this a very interesting result.
However, the introduction of anti D3-branes is not the only way to uplift.
It may also be performed through $F$-terms~\cite{Blumenhagen:2009gk} or $D$-terms on supersymmetric branes~\cite{Burgess:2003ic,Jockers:2005zy,Cremades:2007ig}.

In cases where the supersymmetric contribution to kinetic mixing vanishes, we may yet generate K\"ahler potential corrections yielding kinetic mixing, at much smaller values than (\ref{SUSYKMEstimate}), where the gauge bosons couple to the $F$- and $D$-terms $\bra W^{c}_\alpha\ket =\theta_\alpha D^c, \bra\Xi\ket = \theta^2 F_{\Xi}$ that break supersymmetry via
\begin{align}
\Delta \mathcal{L} \supset& - F_{\mu \nu }^{(a)} F^{(b) \mu \nu} c_{F}^{(1)} \frac{F_{\Xi}}{M^2} \label{SPECIAL:firstorder}\\
& - F_{\mu \nu }^{(a)} F^{(b) \mu \nu}c_F^{(2)} \frac{|F_{\Xi}|^2}{M^4} \label{SPECIAL:secondorderF}\\
&- F_{\mu \nu }^{(a)} F^{(b) \mu \nu} c_D \frac{(D^c)^2}{M^4}, \label{SPECIAL:secondorderD}
\end{align}
where the scale $M$ can be set to the string scale, with the coefficients absorbing the quantitative information. There is no term of $\C{O}(D^c)$ since there is no operator that can generate it: the above are generated by corrections to the K\"ahler potential of the form
\begin{align}
\Delta \mathcal{L} \supset& \int d^4 \theta W^a W^b \frac{\Xi + \ov{\Xi}}{M^2} + c.c.\label{firstorder}\\
&+\int d^4 \theta W^a W^b \frac{{\mathcal{D}}^2 (\ov{\Xi} + \Xi)^2}{M^4} +...\label{secondorder}\\
&+\int d^4 \theta W^a W^b \frac{\ov{W}^c \ov{W}^c}{M^4}.\label{secondorderD}
\end{align}
Note that the above is schematic; the full set of operators involves permuting the insertions of superspace derivatives ${\mathcal{D}}$ etc..
Now, in LARGE volume models, of the gravity-mediated contributions only the dilaton and K\"ahler moduli obtain $F$-terms, and $D$-terms are initially considered to vanish, but then are corrected by uplifting - and, of course, the Higgs vev gives a $D$-term to the hypercharge. Following \cite{Blumenhagen:2009gk}, we shall assume that there are three classes of cycles with corresponding K\"ahler moduli: ``large'' cycles $\tau_b$, ``small'' cycles (which intersect no others and are wrapped by a D3-brane instanton $\tau_s$, and collapsed ``matter'' cycles $\tau_a$.
We will also denote hidden ``collapsed'' cycles $\tau_h$. The $F$ terms are \cite{Blumenhagen:2009gk} parametrised in terms of numbers $\gamma, \xi$ and the constant term in the superpotential after complex structure modulus stabilisation, $W_0$:
\begin{align}
F_S &\approx \frac{3}{2\sqrt{2}} \gamma \frac{\xi}{g_s^2} \frac{W_0}{\C{V}^2}, \nonumber \\
F_{a,h} &=0,  \\
F_b &\approx -2\tau_b M_{3/2} - \frac{3}{8\sqrt{2}} \frac{\tau_b}{a\tau_s} \frac{W_0}{\C{V}^2},\nonumber
\end{align}
where $F_S$ is the dilaton $F$-term, and
\begin{equation}
M_{3/2} = \frac{g_s^{2}|W_0|}{\sqrt{2}\C{V}} M_P.
\end{equation}
The $D$ terms from uplifting are very model-dependent (they correspond to blow-up moduli on branes), but the hypercharge $D$-term in the supergravity basis is
\beq
D_Y = - \frac{1}{2} g_Y k_{H\ov{H}} v^2 \cos 2\beta ,
\eeq
where $k_{H\ov{H}}$ is the Higgs' K\"ahler metric (in the physical basis this is factored out), $v \sin \beta = \sqrt{2} \langle H_u \rangle, v \cos \beta
= \sqrt{2} \langle H_d \rangle$, $v \simeq 246$~GeV is the vev of the Standard Model Higgs.

We would now like to estimate the coefficients appearing in equations \eqref{SPECIAL:firstorder},\eqref{SPECIAL:secondorderF},\eqref{SPECIAL:secondorderD}, and
determine whether they might vanish. Firstly, we should note that if the $F$-term spurions are the K\"ahler moduli $\tau_b, S$ then
they will have axionic couplings to gauge fields; the expressions should then be a function of terms like $\Xi + \ov{Xi} + \Pi^i V_i$.
However, replacing $\Xi$ in Eq.~\eqref{firstorder} with $\Pi^{i}V_{i}$ yields a term with three gauge fields. This is excluded by Furry's theorem.
Consequently, terms of the form \eqref{SPECIAL:firstorder}, \eqref{firstorder} are excluded, too.
In addition, from the CFT point of view it is possible to see that such an amplitude is forbidden by worldsheet
charge conservation. We can then turn to CFT computations on toroidal backgrounds to estimate the magnitude of the coefficients.

Let us now turn to the contribution~\eqref{SPECIAL:secondorderF},\eqref{secondorder}.
Since we have already argued that the visible sector should always mix supersymmetrically with a hyperweak gauge group, we shall focus on interactions
between branes at fixed points of orbifolds. A calculation involves the scattering of two gauge bosons and the appropriate $D$ or $F$ term vertex operators,
and includes contributions from $N=4$ and $N=2$ sectors. The presence of $N=2$ sectors implies shared cycles, which as we also argued yields
supersymmetric mixing. So we need only consider the $N=4$ sectors, which are identical to the interaction between D3-branes except for Chan-Paton factors.
However, since the operators corresponding to the K\"ahler modulus and dilaton $F$-terms do not carry Chan-Paton factors, and the $N=4$ sector amplitude is
independent of the fluxes on the branes, we conclude that the contributions to (\ref{SPECIAL:secondorderF}) cancel.

Turning now to the $D$-terms, we shall restrict our attention to the contributions to (\ref{SPECIAL:secondorderD}) of\footnote{If the visible U(1) arises from a traceless generator of a GUT group terms linear in $W$ vanish. This forbids terms
where the $D$-terms arise from a (broken) non-abelian U(1). This is the case we consider here because otherwise we would most likely already
have SUSY kinetic mixing which would dominate.}
\begin{align}
\Delta \mathcal{L} \supset& \int d^4 \theta W W' \frac{\ov{W}' \ov{W}'}{M^4} + (W W' \frac{W' W'}{M^4} + c.c.)\nonumber \\
&+\int d^4 \theta W W' \frac{\ov{W} \ov{W}'}{M^4} + (W W' \frac{W W'}{M^4} + c.c.) \\
&+\int d^4 \theta W' W \frac{\ov{W} \ov{W}}{M^4} + (W' W \frac{W W}{M^4} + c.c.), \nonumber
\end{align}
where $W$ is the hypercharge gauge superfield and $W'$ is the hidden U(1). Contributions to these operators are shown in Fig.~\ref{dterms} both in
field and in string theory. A field theoretical evaluation of Fig.~\ref{dfield} is given in Appendix \ref{susyfield}.
Note that we consider only the case with two different U(1)s in the operator. In string theory a particle can be charged only under
two U(1)s. Therefore, operators with more than two different U(1)s can appear only at two or more loops.

To determine the mass scale $M$ appearing in the above we can attempt a calculation for toroidal orientifolds.
One method to obtain the coefficient is to calculate the scattering amplitude of four gauge bosons; conveniently this has been performed  in \cite{Bianchi:2006nf}.
The result for the $N=4$ sector, as discussed in Appendix \ref{APPENDIX:TOROIDAL}, is
\beq
\frac{1}{M^4} \sim \frac{1}{4\pi^5 M_s^4} \frac{1}{\C{V}^{2/3}},
\eeq
up to an order one constant. The key is that the $N=4$ sectors will always be present since they are mediated by untwisted modes, linking two hidden sectors, and provides a lower bound upon the mixing.

\begin{center}\begin{figure}
\begin{center}
\subfigure[]{
\includegraphics[angle=-90,width=6cm]{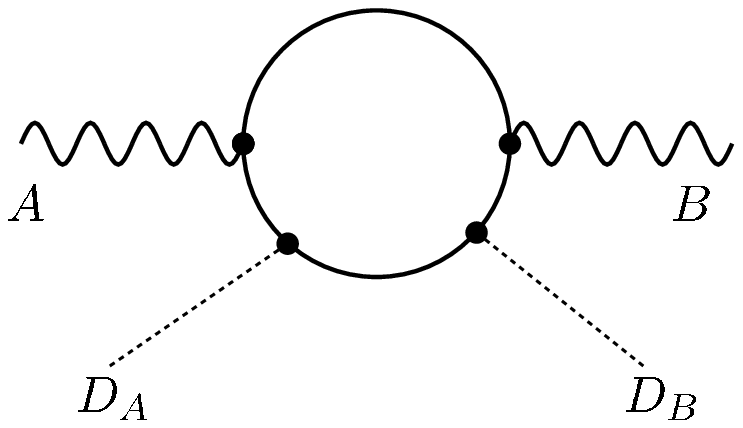}
\label{dfield}}
\hspace*{1cm}
\subfigure[]{
\includegraphics[angle=-90,width=3.5cm]{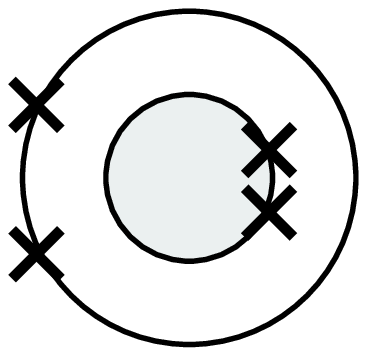}
\label{dstring}}
\end{center}
\caption{Contributions to the kinetic mixing between the gauge groups $A$ and $B$, in presence of SUSY breaking $D$-terms.
The left hand side shows the field theoretic contribution
whereas the right hand side shows the corresponding string theory diagram.
}\label{dterms}
\end{figure}\end{center}

Of course, the above amplitudes are also proportional to the Chan-Paton factors, and thus certain amplitudes may still vanish; this is appropriate for the $N=4$ sectors which are insensitive to the fluxes at the singularity. Therefore for these sectors, the operators $W (W')^3$ will vanish, but we expect $W^2 (W')^2, W^3 (W')$ to be present.

We thus finally have the estimate of kinetic mixing induced by the $D$-term for the hypercharge, given by the operator $ W^3 (W')$, of
\beq
\chi_{hY} (v) \sim  \frac{g_Y^2}{4} \frac{1}{\C{V}^{2/3}} \frac{g_{\rm h} g_Y}{4\pi^5}\left(\frac{v}{M_s}\right)^4 \cos^2 2 \beta ,
\eeq
where $g_{\rm h}$ is the gauge coupling of the hidden U(1), $v\sin\beta,v\cos\beta$ are the up and down-type Higgs vevs, respectively.
This can lead to extremely small values.
For instance, using $\tan \beta \sim 10$ we have $\chi \sim 10^{-59}$ for $M_s \sim 10^{15}$~GeV  and $\chi \sim 10^{-27}$ for $M_s \sim 1$~TeV.

If we consider that the hidden U(1) has a $D$-term generated by fluxes on the collapsed brane \cite{Burgess:2003ic,Jockers:2005zy,Cremades:2007ig},
rather than electroweak symmetry breaking, then the magnitude of the $D$-term may be much larger.
However, due to the vanishing of the operator $W (W')^3$ the contributions will still be very small - the relevant term becomes $W^2 (W')^2$.
For a hidden $D$-term $\C{O}(M_s)$ we obtain mixings of $10^{-33}$ and $10^{-25}$ for $10^{15}$~GeV and TeV  strings, respectively.

We have summarised the predictions from the SUSY breaking contributions to the kinetic mixing identified in this section in Fig.~\ref{Fig:chivsms_susybroken}. We find that the soft breaking contributions associated with $D$-terms are very small, irrespective of the string scale.
However, as emphasised already at
the beginning of this subsection, this does not mean that they are uninteresting. Such tiny values are extremely welcome for interpretations of cosmic ray data in terms of decaying hidden photino dark matter (item 5 in the Introduction).
On the other hand, kinetic mixing with $\ov{\rm D3}$-branes can be substantially larger at sufficiently large string scales.

\section{St\"uckelberg masses}\label{Stueck}

Since we are interested in massive hidden photons let us turn to the mass terms.
For U(1) gauge fields we can have St\"uckelberg as well as Higgs masses. In this section we discuss
St\"uckelberg masses of U(1)s from D7-branes and in the next Sect.~\ref{Higgs} we will turn to masses arising from a Higgs mechanism.

In the 4D effective theory St\"uckelberg masses arise from terms
of the form (see~\cite{Buican:2006sn,Conlon:2008wa,Plauschinn:2008yd}),
\begin{equation}
{\mathcal{L}} \supset -G_{ij} H^{i,\,\mu\nu\rho}H^{j}_{\mu\nu\rho}- \frac{1}{4 g_i^2}F^{i,\,\mu\nu}F^i_{\mu\nu}- \hat{\Pi}_{ij} E^i_{2}\wedge F^j_{2},
\label{SchematicStueck}\end{equation}
where the first two -- kinetic -- terms arise from dimensional reduction of the DBI action and the
third term from dimensional reduction of the Chern-Simons action
in Eq.~\eqref{EFFECTIVE:DBI}, respectively. In particular,
the first term  is the kinetic term for the 2-forms $E^{i}_{2}$, $dE^{i}_{2}\wedge\star_{4}dE^{j}_{2}=\sqrt{g}H^{i,\,\mu\nu\rho}H^{j}_{\mu\nu\rho}$ and $G_{ij}$ is a
non-canonical (generically non-diagonal) K\"ahler potential for the forms, $\hat{\Pi}_{ij}$ are dimension-mass couplings; these both depend upon the details of the Calabi-Yau moduli space, and whether $E_2$ descends from a four-form or six-form (respectively two and four on the Calabi-Yau); $E_2^i = E_2^\alpha(x) \omega_\alpha, E_{2,\,a} \tilde{\omega}^a$ respectively. For the definitions of $\omega_\alpha, \tilde{\omega}^a$ and further details reviewing Kaluza-Klein reduction of 10D IIB supergravity and the
D-brane action we
refer the reader to Appendix \ref{APPENDIX:EFFECTIVE}.
These couplings are given respectively in terms of
\begin{eqnarray}
p_{aD_1 A} &=& \frac{1}{2\pi}\int_{\alpha_A \subset D_1}  F_a , \nonumber \\
b_{D_1 A} &=& \frac{1}{l_s^2}\int_{\alpha_A \subset D_1} B ,
\end{eqnarray}
and
\begin{align}
\Pi_{\alpha}^{D_1 A} =& \frac{1}{l_s^2}\int_{\alpha_A \subset D_1}  \omega_{\alpha}, \nonumber \\
\tilde{\Pi}^{c D_1} =& \frac{1}{l_s^4}\int_{D_1} \tilde{\omega}^c.
\end{align}
Here, $D_i$ denote four-cycles, $\alpha_A$ are two-cycles.
The latter are (possibly zero) integers, or in the case of $b_{D_1 A}$, half-integers
(since we only require the integrals over positive cycles). $r_{a D_1} $ is the D7 brane charge on cycle $D_1$ due to stack $a$.

The kinetic terms $K_{ij}$ are given in terms of the metrics (see also (\ref{EFFECTIVE:METRICS})) in the space of the 2-forms on the Calabi-Yau,
\begin{eqnarray}
G_{\alpha \beta}\!\! &=&\!\! \frac{1}{l_s^6} \int \omega_{\alpha} \wedge \star \omega_{\beta} = -K_{\alpha \beta \gamma} t^\gamma + \frac{\tau_\alpha \tau_\beta}{\mathcal{V}}
\sim \mathcal{V}^{1/3},
\\\nonumber
G_{ab}\!\!&=&\!\! - K_{ab \gamma} t^\gamma \sim  \mathcal{V}^{1/3}.
\label{MetricBehaviour}\end{eqnarray}
As usual the metrics with both indices raised or lowered are the inverse of each other,
\beq
G^{\alpha \beta} \sim G^{ab} \sim \mathcal{V}^{-1/3}.
\eeq
These scalings are for bulk cycles, corresponding to non-anomalous U(1)s; for anomalous U(1)s the dual
cycles are vanishing and the masses generated are at the string scale. For two-forms on the Calabi-Yau the kinetic term $K_{ij}$ is given by $G_{\alpha \beta}$, while for four-forms we need $G^{ab}$.

Collecting everything together the St\"uckelberg masses are given by~\cite{Buican:2006sn,Conlon:2008wa,Plauschinn:2008yd}
\begin{eqnarray}
\label{Stmass}
\lefteqn{
m_{{\rm St}\,ab}^2 = \frac{g_a g_b}{4\pi }\,  M_s^2  }
\\ \nonumber && \times
\bigg[  G_{cd} \tilde{\Pi}^{c D_1}  \tilde{\Pi}^{d D_2} r_{a D_1} r_{b D_2} +   G^{\alpha \beta} \Pi_{\alpha}^{D_1 A} \Pi_{\beta}^{D_2 B} (p_{aD_1 A} - r_{a D_1}  b_{D_1 A}) (p_{bD_2 B} - r_{b D_2} b_{D_2 B}) \bigg].
\end{eqnarray}

We see that the first and second term in the St\"uckelberg mass squared matrix~\eqref{Stmass}
scale with different powers of $\mathcal{V}$: the first contribution will generically make a larger contribution to the U(1) masses than the second.
In fact, for the case that the field labeled by $a$ is residing
on a brane wrapping a vanishing cycle and that the field labeled by $b$ is residing on a hyperweak cycle, we may estimate the first contribution as
\begin{equation}
m^2_{{\rm St}\,(1)}
=  \frac{g_s}{2} M_s^2 \left( \begin{array}{ll}  \sim \mathcal{V}^{1/3} & \sim1 \\
\sim1 & \sim\mathcal{V}^{-1/3} \end{array} \right) ,
\label{Stmass1_est}
\end{equation}
while the second contribution is expected to be of order
\begin{equation}
m^2_{{\rm St}\,(2)}
=  \frac{g_s}{2} M_s^2 \left( \begin{array}{ll}
\sim \mathcal{V}^{-1/3}  & \sim \mathcal{V}^{-2/3} \\
\sim \mathcal{V}^{-2/3} & \sim \mathcal{V}^{-1} \end{array} \right) .
\label{Stmass2_est}
\end{equation}
If present the first term dominates. However,
in LARGE volume scenarios it is often assumed that the Betti number $b^2_-$ is zero.
Then the $\tilde{\Pi}$ in Eq.~\eqref{Stmass} vanish and the latter mass squared matrix \eqref{Stmass2_est} becomes relevant.

Let us see how a mass matrix of the structure Eqs.~\eqref{Stmass1_est}, \eqref{Stmass2_est} can obey the phenomenological constraints.
If we want the (non-hyperweak) gauge group $a$ with gauge coupling $g_{a}$ to correspond to the Standard Model hypercharge it is clear that the
$m^2_{{\rm St} aa}$ entry in the St\"uckelberg mass squared matrix~\eqref{Stmass1_est}, \eqref{Stmass2_est} is
much too large. For all phenomenologically valid values of the string scale ($M_s\gtrsim $~TeV) it would violate bounds
on the photon mass ($m^2_{{\rm St} aa}$ is typically larger than the $Z$ boson mass).
Therefore this term must vanish, as must the non-diagonal terms. An alternative would be to identify the photon with a massless linear combination
of the two gauge groups. However, one can easily check that with a hierarchical mass structure as in Eqs.~\eqref{Stmass1_est}, \eqref{Stmass2_est} the massless
U(1) would have a gauge coupling nearly as small as the hyperweak $g_{b}$.
Therefore the only realistic option is that only $m^{2}_{{\rm St} bb}$ is non-vanishing.

\begin{center}\begin{figure}
\centerline{
\includegraphics[angle=-90,width=8cm]{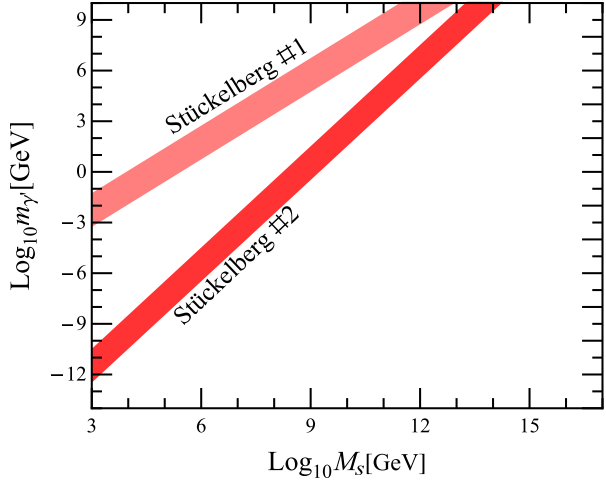}}
\caption{
St\"uckelberg mass term of the hidden U(1) on a hyperweak, bulk D7 brane wrapping a 4-cycle from the estimates in
\eqref{stueckel1} ($\# 1$, light red) and \eqref{stueckel2} ($\# 2$, dark red).
}
\label{Fig:Stmass}
\end{figure}\end{center}

The simplest way to ensure  a St\"uckelberg mass matrix of the phenomenologically viable
form (\ref{req_val}) is to assume that the $\tilde{\Pi}^{cD_{i}}$, $\Pi_{\alpha}^{D_i A}$ are zero for
every cycle for the photon (this is possible if it wraps cycles that are non-trivial on the singularity but trivial in the Calabi-Yau) but non-zero
for the hidden U(1). In this case, only the $m^2_{{\rm St} bb}$ component of the St\"uckelberg mass squared mixing matrix is non-zero.
If the $\tilde{\Pi}$ for the hidden photon are non-vanishing we have
\begin{equation}
\label{stueckel1}
m_{\gamma^\prime}^2 \simeq \frac{g_s}{2}  \left( \frac{4\pi}{g_s^2} \frac{M_s^2}{M_P^2}\right)^{\frac{1}{3}} M_s^2.
\end{equation}
If all the $\tilde{\Pi}$ vanish the first mass matrix~\eqref{Stmass1_est} vanishes completely. Then the hidden photon mass is much smaller and given by the
$m^2_{{\rm St} bb}$ component of Eq.~\eqref{Stmass2_est},
\begin{equation}
\label{stueckel2}
m_{\gamma^\prime}^2 \simeq \frac{g_s}{2}  \left( \frac{4\pi}{g_s^2} \frac{M_s^2}{M_P^2}\right) M_s^2.
\end{equation}
These estimates are plotted in Fig.~\ref{Fig:Stmass}. For low string scales the mass of the hidden photon can be as small as $\sim 1\,{\rm meV}$.

\section{Hidden photon masses from a Higgs mechanism}\label{Higgs}

In the last section we have discussed the possibility that the mass for the
hidden photon arises from a St\"uckelberg mechanism.
The alternative is that it arises via a Higgs
mechanism. Since for a high string scale the St\"uckelberg masses are large it is therefore interesting to
examine alternative means of generating (potentially observable) small masses.
In the following we will discuss how a small symmetry breaking scale can
arise for hidden U(1) gauge groups in LARGE volume scenarios.
In particular, we may also consider the case that the hidden U(1) is not hyperweak: there
may be hidden U(1)s at a hidden singularity with properties similar to those of the MSSM sector.
The two natural values that we shall take for our hidden gauge couplings $g_{\rm h}$ are then the electroweak value, $g_Y$, and
the hyperweak one, $g_Y\C{V}^{-1/3}$.

Breaking the U(1) with a hidden Higgs mechanism leads to some interesting features such as, e.g.,
the existence of a hidden Higgs which can be phenomenologically
interesting. We will discuss this and the constraints on the hidden Higgs from the underlying SUSY in the next subsection.
Then we will estimate the possible range of soft masses that will set the scale for the hidden Higgs vev. Finally we will look at some non-minimal models.

\subsection{General features of a hidden Higgs mechanism}\label{generalfeatures}
\subsubsection*{Phenomenological features}
\begin{figure}[t]
\centerline{\includegraphics[angle=-90,width=10cm]{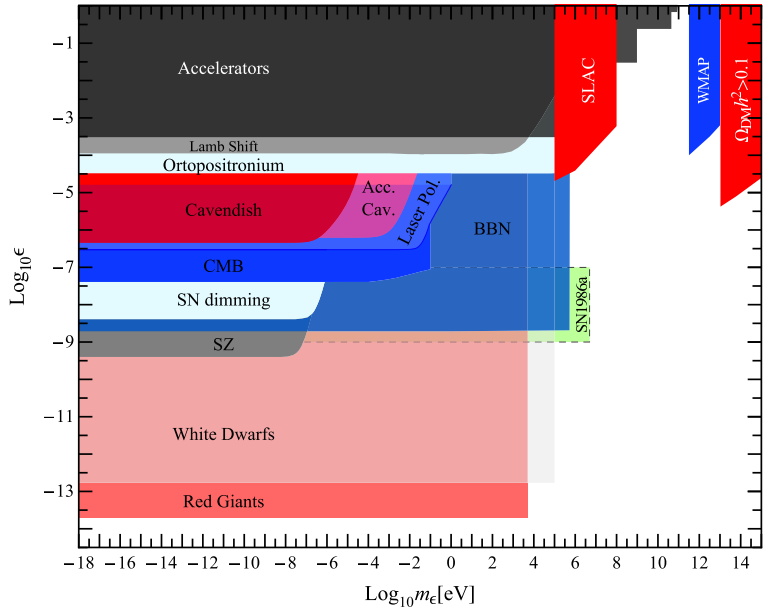}}
\caption{
Constraints on the existence of minicharged particles with electrical charge (in units of the
elementary electric charge $e$) $\epsilon$ and
mass $m_\epsilon$~\cite{Ahlers:2007qf,Davidson:1993sj,Davidson:2000hf,Gluck:2007ia,Badertscher:2006fm,Gies:2006hv,Melchiorri:2007sq,Burrage,Ahlers:2009kh,Jaeckel:2009dh,Dubovsky:2003yn}.
The bounds are usually quoted for models without hidden photons, but the most relevant are also valid where the minicharge arises from kinetic mixing and the hidden photon mass is smaller or comparable to $m_\epsilon$ (The WMAP bound of \cite{Dubovsky:2003yn} is an exception.).
}\label{mcpbounds}
\end{figure}

\begin{figure}[t]
\centerline{\includegraphics[angle=-90,width=10cm]{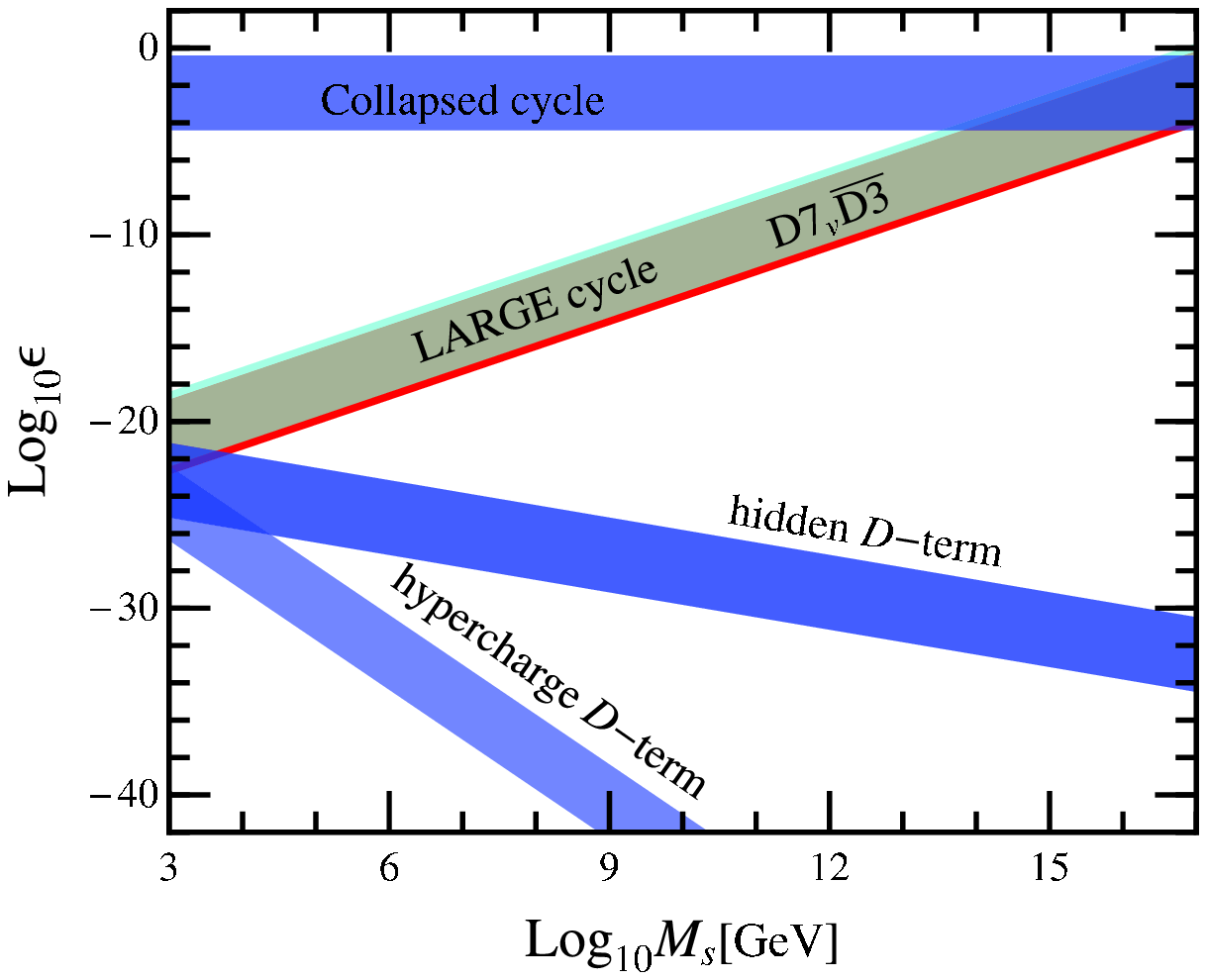}}
\caption{
Natural expectations for the minicharge $\epsilon$ of hidden matter charged under a hidden U(1) mixing kinetically with the electromagnetic U(1). The labels refer to the nature of the hidden U(1) as hyperweak (red), $\overline{\rm D3}$ (turquoise) and collapsed cycle (blue). Note that the former two overlap which each other.
Also shown are corrections to the latter case from SUSY breaking through $D$-terms in the electroweak sector or in the hidden sector.
}\label{mcps}
\end{figure}

In considering a hidden Higgs mechanism as a means of naturally obtaining light U(1) masses, we should also examine the phenomenology of these new fields.
As discussed in \cite{Ahlers:2008qc} the hidden Higgs would in many situations behave like a minicharged particle with
(fractional) electric charge
$\epsilon_{H}\sim \frac{g_{\rm h}}{g_{\rm vis}}\chi$ on which strong bounds exist $\epsilon_{H}\lesssim 10^{-14}$ if the mass is small, cf. Fig.~\ref{mcpbounds}.
However, this bound is not quite as restricting as it seems at first glance.
First, if the hidden Higgs mass is $\gtrsim {\rm few}\,{\rm MeV}$ these bounds weaken
dramatically to $\epsilon_{H}\lesssim 10^{-5}$. Secondly we note that
in hyperweak scenarios
\begin{equation}
\epsilon_{H}\sim \frac{g_{\rm h}}{g_{\rm vis}}\chi\ll \chi.
\end{equation}
Hence, the constraint on $\epsilon_{H}\lesssim 10^{-14}$ only leads to a constraint $\chi\lesssim 10^{-8}$ for a hidden gauge coupling
$g_{\rm h}\sim 10^{-6}$. Therefore, in hyperweak models the constraints on a minicharged Higgs are only relevant for models
with a string scale $M_{s}\gtrsim 10^{9}\,{\rm GeV}$, see Fig.~\ref{mcps}.
From a phenomenological point of view: in hyperweak scenarios searching for a massive hidden
photon is a competitive and sometimes even more sensitive
probe of the extra gauge group than searching for a minicharged Higgs or other minicharged hidden matter.

\subsubsection*{Higgs vev and mass in SUSY models}
In a setup with hyperweak interactions, it is naively straightforward to obtain a light U(1):
\begin{equation}
\label{smallmass}
m_{\gamma^{\prime}}=\sqrt{2}g_{\rm h}\langle \higgs\rangle\ll \langle \higgs\rangle,
\end{equation}
where $H_{\rm h}$ is a hidden sector Higgs.

Using the hyperweak gauge coupling $g_{\rm h}\sim 10^{-10}$ for a low string scale $M_{s}\sim {\rm TeV}$ and a vev for the hidden Higgs
of the order of $\sim100\,{\rm GeV}$ this immediately suggests a very interesting mass in
the eV regime and a kinetic mixing $\chi\sim 10^{-12}$.

However, since we are dealing with a supersymmetric theory we have to be slightly more careful.
The reason is the following. Typically,
\begin{equation}
\label{natural}
\langle \higgs \rangle\sim\sqrt{\frac{\mu^2_{\rm h}}{\lambda_{\rm h}}},\quad{\rm and}\quad m_{\higgs}\sim \mu_{\rm h},
\end{equation}
where $\mu^2_{\rm h}$ is the negative mass-squared responsible for the Higgs vev, $\lambda_{\rm h}$ is the quartic hidden Higgs coupling and $m_{\higgs}$ is
the (physical) mass of the hidden Higgs.
Typically naturalness arguments are based on an argument that $\lambda_{\rm h} \sim 1$ and therefore $\langle\higgs\rangle\sim\mu_{\rm h}\sim m_{\higgs}$. But,
this naive argument fails in supersymmetric models with hyperweak gauge interactions because
\begin{equation}
\lambda_{\rm h}\sim g^{2}_{\rm h}\ll 1.
\end{equation}
Using this and the fact that we expect the negative mass-squared to be roughly of the same order as the soft supersymmetry breaking masses in the hidden sector we get
our generic expectation,
\begin{equation}
\label{soft}
m_{\gamma^{\prime}}\sim m_{\higgs}\sim m^{\rm hid}_{\rm soft}.
\end{equation}
In particular, we typically do not expect a hierarchy between the hidden Higgs mass and the hidden photon mass.

Let us confirm this expectation by looking at the minimal model of one hidden Higgs pair $H_1, H_2$ charged only
under the hidden U(1) with opposite charges $\pm 1$, since
softly broken supersymmetry dictates that the potential is
\beq
V = m_1^2 |H_1|^2 + m_2^2 |H_2|^2 + m_3^2 (H_1 H_2 + c.c) + \frac{1}{2}  (\xi_{\rm h} + g_{\rm h}|H_1|^2 - g_{\rm h}|H_2|^2)^2,
\label{MinmalHiggsV}
\eeq
where $\xi_{\rm h}$ is a Fayet-Iliopolous term. If $\xi_{\rm h} > m_2^2 > 0, m_1^2 >0, m_3^2 = 0$ then the symmetry
breaking is as in \cite{Morrissey:2009ur,Chun:2008by,Cheung:2009qd,Baumgart:2009tn,Cui:2009xq} and we find $m_{\gamma'}^2 = m_{H_2}^2 = 2 (g_{\rm h}\xi_{\rm h} - m_2^2)$.
As expected this allows no hierarchy between the gauge boson and
Higgs masses, so we may not obtain lighter hidden gauge bosons.

If we assume $m_1 \sim m_2 \sim m_3\sim m_{\rm soft}$ then the Higgs masses
are $\sim g_{\rm h} \sqrt{\xi_{\rm h} \pm m^{\rm hid}_{\rm soft}} \sim m_{\gamma'}$.
If we set $\xi_{\rm h} =0$ or equivalently relatively increase the masses $m_i$, then we find a very similar situation;
the lightest Higgs is naturally of similar mass to the heavy gauge bosons.
Schematically we find
$\langle \higgs \rangle^{2} \sim |m_i^2|/g^{2}_{\rm h}$, $m_{\gamma'} \sim g_{\rm h}
\langle \higgs \rangle \sim \sqrt{|m_i^2|} \sim m_{\higgs}$.
We cannot have simultaneously a light hidden gauge boson and comparatively heavy Higgs without a substantial degree of fine tuning.
This is analogous to the situation in the MSSM. Clearly this is a result of supersymmetry dictating the size of the quartic coupling for this minimal situation.

However, the LARGE volume scenario need not give us the minimal scenario. In subsection \ref{stringscenarios} we shall consider what types of
Higgs fields may appear in LARGE volume compactifications and how they might turn out to be heavier than the hidden photon.

Finally let us note, that the case when hidden Higgs does not acquire a vacuum expectation value is interesting in itself.
In this case the hidden photon is massless (unless it gets a mass from the St\"uckelberg mechanism) and the hidden Higgs
is nothing but a charged hidden matter particle. It then behaves as a minicharged particle with
\begin{equation}
\epsilon\sim \frac{g_{\rm h}}{g_{\rm vis}}\chi\sim \frac{1}{16\pi^2}g^{2}_{\rm h}\quad{\rm and}\quad m_{\epsilon}\sim m^{\rm hid}_{\rm soft}.
\end{equation}
If the gauge coupling $g_{\rm h}$ is hyperweak this particle has an interesting phenomenology as it behaves in many
situations as if the hidden photon is not present (cf.~\cite{Burrage}).

\subsection{(Small) soft masses}
In the previous subsection we have seen that the mass scale of the hidden photon is typically given by the soft masses $m^{\rm hid}_{\rm soft}$
arising from supersymmetry breaking in the hidden sector.
Let us now turn to estimating the size of these soft masses in various scenarios of supersymmetry breaking.

\subsubsection*{Gauge mediation}

Let us start with gauge mediated supersymmetry breaking.
In a simple scenario of this type,
\begin{equation}
m^{\rm vis}_{{\rm soft}}\sim \frac{g^{2}_{\rm vis}}{16\pi^2} \frac{M^{2}_{\cancel{\rm SUSY}}}{M_{\rm mess}},
\end{equation}
where $m^{\rm vis}_{{\rm soft}}$ is the typical scale of soft supersymmetry breaking masses\footnote{In more general gauge mediation
scenarios this formula can be modified (see, e.g.,~\cite{Shihextraordinary,Meade:2008wd}).
For example the gaugino, sfermion and Higgs
masses do not necessarily have to be of the same order~\cite{Izawa:1997gs,Kitano:2006xg,Csaki:2006wi,Abel:2007jx,Abel:2007nr,Abel:2008gv,Abel:2009ze}} in the visible, i.e. electroweak,
sector and $M^{2}_{\cancel{\rm SUSY}}$
is the supersymmetry breaking in a sequestered supersymmetry breaking sector.
Moreover, $M_{\rm mess}$ is the mass of the messengers.
Naturally we expect that $m^{\rm vis}_{{\rm soft}}$ sets the scale for electroweak symmetry breaking (with possibly
an additional factor $\sim 1/(16\pi^2)$).

We can now apply the same line of reasoning to the hidden sector. In principle we have two possibilities:
\begin{itemize}
\item[(a)]{} The hidden sector couples directly (via its hyperweak interaction) to the sequestered SUSY breaking sector,
\begin{equation}
m^{\rm hid}_{\rm soft}\sim \frac{g^2_{\rm h}}{16\pi^2} \frac{M^{2}_{\cancel{\rm SUSY}}}{M_{\rm mess}}
\sim \frac{g_{\rm h}}{g_{\rm vis}}m^{\rm vis}_{{\rm soft}}.
\end{equation}
This assumes the most conservative choice of messenger properties; that the supersymmetry breaking scale and messenger masses are identical to those of the visible sector. Of course, there is no a priori reason for this to be true, and we can therefore envisage that there is a very large range of hidden soft masses generated this way.
\item[(b)]{} The hidden sector couples only indirectly via the kinetic mixing with the electromagnetic (hypercharge) U(1) to the sequestered
SUSY breaking sector. This is the case considered in \cite{Morrissey:2009ur,Chun:2008by,Cheung:2009qd,Baumgart:2009tn,Cui:2009xq}, where it transpires that the ``semi-direct mediation'' via the $D$-term of the hypercharge
induces an effective Fayet-Iliopoulos term for the hidden gauge group $\xi_{\rm h}$ of
\beq
\xi_{\rm h} = \chi \langle D_Y  \rangle
\eeq
and thus generate masses of magnitude (including some tachyonic)
\beq
(m^{\rm hid}_{{\rm soft}})^2 = Q_{\rm h} g_{\rm h} \chi \langle D_Y  \rangle = Q_{\rm h} g_{\rm h} g_Y \chi \frac{1}{8} v^2 \cos 2 \beta \ll
\left(m^{\rm vis}_{{\rm soft}}\right)^2.
\eeq
This dominates over the ``little gauge mediation'' contributions
\begin{equation}
m^{\rm hid}_{{\rm soft}}\sim \frac{g_{\rm h}g_{\rm vis}\chi}{16\pi^2} \frac{M^{2}_{\cancel{\rm SUSY}}}{M_{\rm mess}}
\sim \frac{g_{\rm h}\chi}{ g_{\rm vis}} m^{\rm vis}_{{\rm soft}}.
\end{equation}
In fact, the scenario of \cite{Morrissey:2009ur,Chun:2008by,Cheung:2009qd,Baumgart:2009tn,Cui:2009xq} which is of particular phenomenological
interest in the context of hidden sector dark matter (see item 3 in the introduction and references therein and the
region labeled ``Unified DM" in Fig.~\ref{Fig:current_limits})
can be easily realised if we locate the hidden sector on a collapsed cycle analagous to the Standard Model cycle (as in Fig.~\ref{scenarios}(b)), so that $g_{\rm h} \sim g^{}_Y$ and we obtain a hidden U(1)
mass of $\C{O} ({\rm GeV})$. The model requires $\chi \sim 10^{-3} - 10^{-4}$, which corresponds to the supersymmetric contribution $g_Y^2/16\pi^2$, and thus from the earlier discussion we conclude that this scenario can be realised
if the hidden and visible sectors have a homologous collapsed two-cycle.
\end{itemize}

\begin{center}\begin{figure}
\begin{center}
\includegraphics[angle=-90,width=7.5cm]{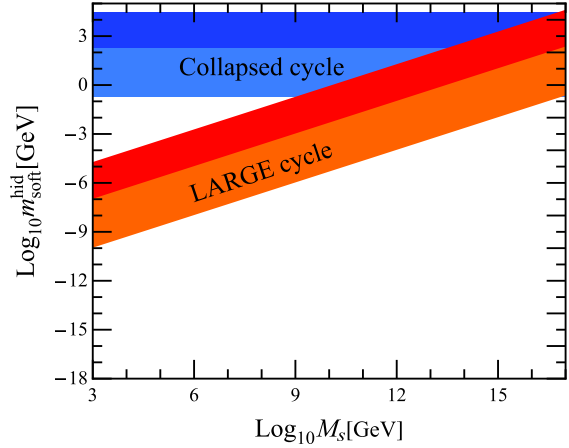}
\label{softgauge}
\end{center}
\caption{
Contributions to soft supersymmetry breaking mass terms in the hidden sector from gauge mediation with
maximally conservative assumptions about the messenger sector.
The blue areas denote soft masses on collapsed cycles and the red areas those on hyperweak branes (LARGE cycles).
The darker color is for the case where the messengers directly couple to the hidden U(1) and the lighter color is for those areas where the coupling
only occurs via kinetic mixing.
}
\label{softmasses}
\end{figure}\end{center}

In Fig.~\ref{softmasses} we show the typical hidden sector soft masses arising via gauge mediation as a function of the string scale, for the minimal choice of messenger sector; less minimal choices would lead to an even wider range of values.
The red areas are for hyperweak hidden sectors whereas the blue areas correspond to hidden sectors on collapsed cycles with gauge
couplings of strength ${\mathcal{O}}(1)$. Darker shades correspond to the mediation via mechanism (a) and lighter shades to mechanism (b).
Overall we note that in gauge mediation scenarios the soft masses in the hidden sector can be many orders of
magnitude smaller than those in the visible sector, in particular if the gauge coupling is hyperweak.

\subsection*{Gravity mediation}

We should however keep in mind that in addition to the gauge mediation contributions there is always an unavoidable contribution
from gravity mediation, but we may not simply take the naive contributions to the soft MSSM and gravitino masses.
This is because while all $F$ terms effectively couple to the gravitino, not
all $F$ terms couple at tree level to the
matter fields. Soft masses for fields with diagonal K\"ahler metric $K_{\alpha \ov{\beta}} = K_\alpha \delta_{\alpha \ov{\beta}}$ are given by
\beq
m_{{\rm soft},\,\alpha}^2 = \left(M_{3/2}^2 + \frac{V_0}{M_P^2}\right) - \ov{F}^{\ov{m}} F^n \partial_{\ov{m}} \partial_n \log K_{\alpha} ,
\eeq
where $V_0$ is the vacuum expectation value of the potential.
For states with generic volume dependence of their K\"ahler metric $K_\alpha = k_\alpha \C{V}^{-2/3+\gamma}$ we have for $\gamma \ne 0$
\beq
m_{{\rm soft},\,\alpha}^2 \sim \frac{3}{2} \gamma M_{3/2}^2.
\label{LeadingMass}
\eeq
In gauge mediation scenarios the SUSY breaking of size $M_{\cancel{\rm SUSY}}$ in the sequestered sector would give a contribution
of size $\sim \frac{M^{2}_{\cancel{\rm SUSY}}}{M_{P}}$ to the gravitino mass. But the gravitino mass could of course be bigger because there could be SUSY breaking
that is not coupled via gauge mediation.

For example, for $\gamma=-1/3$ (as we shall later find corresponding to Higgs localised on large cycles) we
then find a tachyonic mass $m_{\alpha}^2 \sim - \frac{1}{2} M_{3/2}^2 $.

However, as shown in \cite{Blumenhagen:2009gk}, for fields
on a collapsed cycle such as a ``collapsed-collapsed'' or a ``collapsed-large'' Higgs field,
which have K\"ahler metric $K_\alpha \sim \C{V}^{-2/3}$, and assuming $V_0 \sim 0$, the leading terms in the above equation cancel, and we have a
naive contribution from gravity mediation which is much smaller,
\beq
\label{gravsupp1}
m_{{\rm soft},\,\alpha}^2 \sim M_{3/2}^2 \frac{1}{g_s^{3/2} \C{V}}.
\eeq
However, in \cite{Blumenhagen:2009gk} it was conjectured that the K\"ahler metric for chiral matter at collapsed cycles may actually be
\beq
\label{gravsupp2}
K_{\alpha} = e^{K/3},
\eeq
where $K$ is the K\"ahler potential of the moduli. This would then cancel the soft masses
from gravity mediation exactly, leaving only higher order corrections. These must then be compared to any
gauge or anomaly mediated contribution.

The conclusion is that the gravity mediated soft masses may be almost independent of the gravitino mass, and that if we take the visible sector soft terms to arise from gravity mediation then this provides a lower bound for any hidden sector terms.
On the other hand, if the visible SUSY breaking arises from gauge mediation the soft masses arising from gravity mediation in the hidden sector could be nearly arbitrarily small.

\subsection{Higgs fields, masses and expectation values in string theory}\label{stringscenarios}
Having discussed our expectations for the scale of the hidden photon and hidden Higgs masses arising in scenarios with hidden Higgses
let us now turn to concrete string realizations. In this context we will also keep an eye on the possibility that we can obtain
larger Higgs masses than the naive expectation $m_{\higgs}\sim m_{\gamma^{\prime}}$.

Hidden Higgs fields, being scalars, must occur at the intersection between two branes, although this includes
intersections with orientifold images. Since the branes wrap four-dimensional divisors on the Calabi-Yau space, they
either overlap (for ``collapsed-collapsed'' brane matter), having a four-dimensional intersection, or their
intersections are two-dimensional curves (Riemann surfaces); this occurs for ``large-large'' and ``large-collapsed'' matter.
Using the knowledge that the chiral wavefunctions are localised at the intersection and that the
Wilsonian Yukawa couplings do not depend upon the K\"ahler moduli, this allows us to estimate the scaling of the K\"ahler potential with the cycle volumes.
For the states on two-dimensional intersections, if we rescale $\tau_h\rightarrow \beta \tau_h$, then since $\tau_h = \tau_h^i \tau_i = \tau_h^i \frac{1}{2} \kappa_{ijk} t^i t^j t^k$ we must rescale all of the two-cycles that comprise $\tau_h$; then the intersection will rescale as $\sqrt{\beta}$.

We shall take the K\"ahler potential for chiral fields to have volume dependence of
\begin{align}
K^{(bh)} &\sim \frac{1}{\C{V}^{2/3}},\nonumber \\
K^{(hh)} &\sim \frac{1}{\C{V}^{2/3}},
\end{align}
where the $K^{(bh)}$ states are analagous to D3-D7 states in a local orbifold model, and thus we have the above dependence \cite{Conlon:2009xf,Conlon:2009kt}.

\subsubsection{Vector pair of Higgs}

An intriguing possibility is that we have a vector-like pair of hidden Higgs fields at the intersection of the large cycle with its orientifold image.
The number of vector-like Higgs fields $\higgs, \higgst$ is counted
by $\min [h^0(C,L_a^{\vee} \otimes L_a^\prime \otimes K_C^{1/2}), h^1(C,L_a^{\vee} \otimes L_a^\prime \otimes K_C^{1/2})]$ where $C$ is the curve over which the brane and image intersect; this is given by the two-cycle $C^i (\omega_i)$, where
\beq
C^i = \int [D_a] \wedge [D_a^\prime] \wedge \omega_i.
\eeq
Then, since we are perturbatively allowed a $\mu$ term, we find the scaling with volume of the K\"ahler potential of such states, since the intersection grows as $\sqrt{\tau_b}$, is
\beq
K^{(bb)} \sim \C{V}^{-1}.
\eeq
This leads to a tachyonic gravity-mediated mass using equation (\ref{LeadingMass}) for the Higgs pair of
\beq
m_{\higgs}^2 \sim - M_{3/2}^2.
\eeq
In such a scenario, we can also consider the possibility that the large cycle is not rigid, allowing an adjoint
singlet $S_{\rm h}$. Since we are not concerned about the hidden U(1) unifying into a larger gauge group, and that the
cycle can be potentially stabilised by supersymmetry breaking effects, it is reasonable to consider this possibility. In such a case,
there will be a holomorphic Yukawa coupling to the vector Higgs pair of
\beq
W \supset \lambda_S S_{\rm h} \higgs \higgst .
\eeq
By dimensional reduction of the Dirac-Born-Infeld action, the adjoint K\"ahler potential scales
as $K^{(S_{\rm h})} \sim \C{V}^{2/3}$, and thus the gravity-mediated mass generated for it, using equation (\ref{LeadingMass}), is
\beq
m_{S_{\rm h}}^2 \sim 2 M_{3/2}^2 .
\eeq
This is large and positive, stabilising the adjoint at zero vacuum expectation value. Then, as shown in \cite{Belanger:2009wf} in the context of global supersymmetry, we have an additional contribution to the physical (i.e. normalised) Higgs quartic coupling of
\beq
\label{quartichiggs}
V \supset \frac{1}{4}\frac{\lambda_S^2}{\C{V}^{2/3}} \higgs^{2} \higgst^2,
\eeq
after integrating out the scalars for $S_{\rm h}$. This can allow an increase in the hidden Higgs mass compared to the naive expectation.
However, note that in Eq.~\eqref{quartichiggs} the coupling scales like $g_{\rm h}^2$
-- so we wouldn't expect an increase by orders of magnitude.
This is because for a very specific value of the (holomorphic and thus K\"ahler modulus independent) $\lambda_S$
the fields are $N=2$ supersymmetric. This occurs when the physical coupling
$\lambda_S e^{K/2} (K^{(S_h)} K^{(bb)} K^{(bb)})^{-1/2} = \sqrt{2}g_{\rm h}$ (assuming
the hidden Higgs field has charge $\pm 1$ under the hidden gauge group).
Since there is no such $N=2$ symmetry for a general Calabi-Yau orientifold, we
generically expect $\lambda_S$ to differ from this value -- possibly allowing for a significant increase of the hidden Higgs mass.

\subsubsection{Chiral Higgs}\label{chiral}

We can also consider the Higgs to be a chiral state, intersecting the hyperweak brane and a hidden cycle.
In string models we find that chiral Higgs, in contrast to a Higgs pair stretching between brane and orientifold image as above,
will always be charged under an additional gauge group (which may be anomalous).
If this group has a coupling $\tilde{g}_{\rm h}$ that is not hyperweak (i.e. if the hidden Higgs is at an
intersection between the hyperweak gauge group and a hidden sector on collapsed cycles - exactly the place where
the gravity mediated masses may be acceptable) then we have an additional quartic coupling contribution and the Higgs
masses will be enhanced relative to their vev. However, if this gauge group is massless,
then we will find a rank one mass matrix
for the two gauge bosons, i.e. one linear combination will remain massless, as in the MSSM.
The natural remedy for this is that they are charged under an additional hidden broken U(1) that has a mass from the St\"uckelberg mechanism.
For an appropriate St\"uckelberg mass, we then expect the hidden Higgs breaking to give
$m_{\gamma'}^2 \sim |(m^{{\rm hid}}_{\rm soft})^{2}|g^2_{\rm h}/\tilde{g}^2_{\rm h} \ll m^{2}_{\higgs} \sim |(m^{{\rm hid}}_{\rm soft})^{2}|$.
However, we must be careful since the St\"uckelberg couplings, being supersymmetric, affect the $D$-terms.

We saw in section \ref{Stueck} that St\"uckelberg masses are generated by the coupling of space-time two-forms to the gauge fields; these two-forms are associated to either orientifold-even two-forms or odd four-forms on the compact space. However, space-time two-forms are dual to scalars, and it is the scalars that appear in the supergravity Lagrangian, as the lowest components of linear multiplets. However, they can be equivalently written in terms of chiral multiplets, and as this is the form that the moduli are usually written in the literature we shall follow this convention. The couplings $\Pi_{ij}$ \eqref{SchematicStueck} are proportional to the string mass and numerical factors (depending on the magnetic fluxes and cycles wrapped) but not the volume or string coupling. When we dualise (keeping the discussion general so that the superfield is $X_i = x_i + iC_i$, where $C_i$ is the axion and $x_i$ a modulus) we find the K\"ahler potential to quadratic order is
\beq
K \supset \frac{1}{4} K_{ij} (X_i + \ov{X}_i +  2\Pi_{ik} g_k V_k)(X_j + \ov{X}_j  + 2\Pi_{jl} g_l V_l),
\eeq
where $V_i$ are the chiral superfields with couplings $g_i$. The K\"ahler potential $K_{ij} = \frac{1}{2} G^{i j}$, the inverse of $G_{ij}$ in \eqref{SchematicStueck}. We shall take $K_{ij} = k_x \C{V}^{-2/3+\gamma}, \Pi_{ij} = \Pi_x M_s; k_x, \Pi_x \sim \C{O}(1)$. Then the physical mass for the additional hidden $U(1)''$ is given by
\beq
m_{\gamma''}^2 = \tilde{g}^2_{\rm h} M_s^2 k_x \C{V}^{-2/3+\gamma} \Pi_x^2 \sim  \tilde{g}^2_{\rm h} M_s^2 \C{V}^{-2/3+\gamma}.
\eeq
For bulk two-forms we have $\gamma = 1/3$ (c.f. \eqref{MetricBehaviour}); for bulk four-forms we have $\gamma=1$; and for anomalous U(1)s (whose two- or four-forms are localised) we have $\gamma=2/3$. Only two-forms will yield masses suppressed relative to the string scale for branes on collapsed cycles, and so this is the case to which we shall restrict ourselves.

We then have the additional contribution to the potential
\beq
\Delta V =  \frac{1}{2}K_{ii} m_x^2 x_i^2 + \frac{1}{2} \tilde{g}^2_h (K_{ii} \Pi_i M_s x_i + k_{H\ov{H}} (|H_1|^2 - |H_2|^2))^2,
\eeq
where $k_{H\ov{H}}$ is the hidden Higgs K\"ahler metric, assumed to be diagonal and identical for $H_1, H_2$; $m_x$ is the physical mass of the modulus $x_i$, which is then stabilised, modifying the quartic coupling to
\beq
\frac{g^2_{\rm h}}{2} \rightarrow \frac{g^2_{\rm h}}{2} + \frac{\tilde{g}^2_{\rm h}}{2}
\left(\frac{  m_x^2}{ m_x^2 + K_{ii} \tilde{g}^2_{\rm h} M_s^2 \Pi_x^2}\right),
\eeq
where we have rescaled the Higgs fields; the low energy theory thus has an effective potential of
\beq
\tilde{V} = m_1^2 |H_1|^2 + m_2^2 |H_2|^2 + m_3^2 (H_1 H_2 + c.c) + \frac{1}{2} \bigg[g^2_{\rm h} + \tilde{g}^2_{\rm h} \left(\frac{  m_x^2}{ m_x^2 + m_{\gamma''}^2}\right)\bigg] (|H_1|^2 - |H_2|^2)^2 .
\eeq
We can write $g^2_{\rm h} \sim \tilde{g}^2_{\rm h} \C{V}^{-2/3}$, and thus require
\beq
\frac{m_{H_{\rm h}}}{m_{\gamma'}} \sim \C{V}^{1/3}\left(\frac{  m_x^2}{ m_x^2 + m_{\gamma''}^2}\right)^{1/2} \sim \sqrt{g_s/2}\, |W_0| \gg 1,
\eeq
where the latter identity is for two-form generated mass (clearly if the $U(1)''$ is anomalous and has a string scale mass then we cannot use this mechanism) and we have set $\tilde{g}_h^2 = 2\pi g_s$.

Since the bounds on light minicharged particles are significantly
relaxed for $m_{\higgs} \gtrsim {\rm MeV}$, we can use this to constrain the minimum hidden photon mass given a
certain gravitino scale/fine tuning.
From the above, we find
\beq
m_{\gamma'} \gtrsim  \frac{1}{|W_0|} \ \mathrm{MeV},
\eeq
up to order one factors (including $\sqrt{g_s/2}$).
The ``natural'' value for $W_0$ is generally assumed to be $1$; so to have very light hidden photons requires some fine tuning.
For example, to obtain a hidden photon mass of $\sim{\rm eV}$ with a
hidden Higgs mass of $\sim{\rm MeV}$ we would need $W_0 \sim 10^6$; however it would not be unreasonable to expect a hierarchy of $10^2 - 10^4$, allowing hidden photon masses as low as $100$~eV. We thus have a natural mechanism that allows a moderate hierarchy between the hyperweak gauge boson and hidden Higgs masses.

\section{Summary and conclusions}\label{conclusions}

\begin{figure}[ht]
\centerline{\includegraphics[,angle=-90,width=15cm]{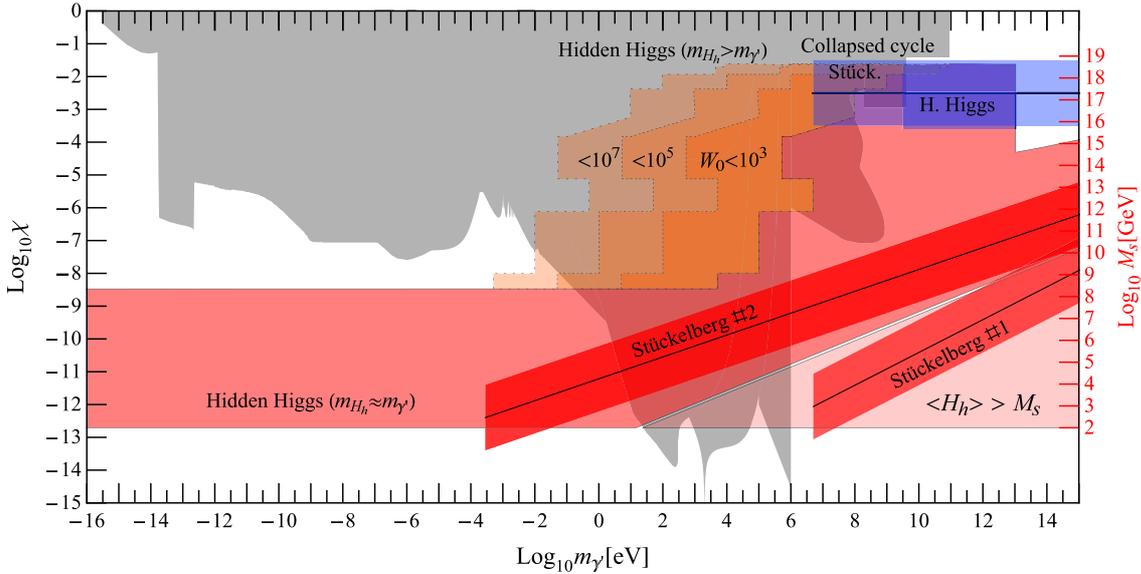}}\vspace{0cm}
\caption{
String theory expectations for the mass and kinetic mixings of massive hidden U(1) gauge bosons kinetically mixing with the ordinary electromagnetic photon.
The red/orange areas denote a situation where the hidden U(1) arises from a hyperweak D7 brane.
In this case the kinetic mixing is directly related (up to an order of magnitude) to the string scale $M_s$, which is shown on the right axis.
The straight bands denote the situations where the hidden photon mass arises from a St\"uckelberg mechanism.
The lighter red area extending to arbitrarily low masses shows the possible masses from a Higgs mechanism.
The lower bound comes from a lower bound to the string scale $M_s >100$ GeV.
In the lower right triangle (lighter red) shows a region where the vacuum expectation value of the hidden Higgs
would be larger than the string scale ($\langle H_{\rm h}\rangle >M_s$).
The upper bound on the kinetic mixing comes from constraints on the phenomenology of the hidden Higgs particle which becomes mini-charged (cf. Sect.~\ref{generalfeatures} and \cite{Ahlers:2008qc}).
All the red areas correspond to the natural case where the hidden Higgs mass is similar to that of the hidden photon
($m_{H_{\rm h}}\approx m_{\gamma'}\approx m_{\rm soft}^{\rm hid}$).
However, in models with a chiral Higgs it is possible to create a hierarchy ($m_{H_{\rm h}}\gg m_{\gamma'}$) by tuning $W_0$, which
make the regions shown in orange phenomenologically allowed.
The blue bands correspond to the case where the hidden photon arises from a collapsed cycle and either gains its
mass through a St\"uckelberg or a hidden Higgs mechanism.
For comparison we have included the current experimental constraints as the grey area (cf. Fig.~\ref{Fig:current_limits} for details).
}\label{stringexpectations}
\end{figure}

Extra U(1) gauge bosons kinetically mixing with the electromagnetic (or hypercharge) U(1) may provide us with a unique window into hidden sector physics.
Moreover, they could play a role in a number of observed phenomena possibly connected to dark matter
(cf. items 2-5 in the Introduction and the regions labeled ``Lukewarm DM",
``Unified DM" and ``Hidden Photino DM" in Fig.~\ref{Fig:current_limits}).
In this paper we have investigated what masses and sizes of kinetic mixings are expected in string theory, more precisely
in LARGE volume compactifications. Our findings are summarized in Fig.~\ref{stringexpectations}.

\begin{figure}[t]
\centerline{\includegraphics[angle=-90,width=15cm]{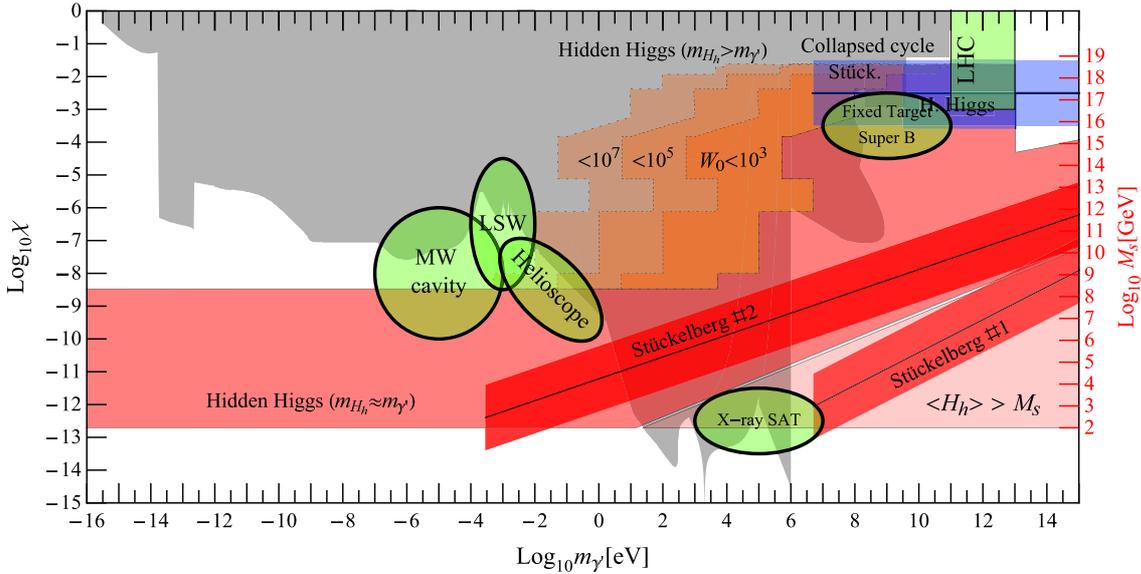}}
\caption{
As in Fig.~\ref{stringexpectations} but in addition we show some interesting future probes of massive hidden U(1)s
like: light-shining-through-a-wall (LSW) experiments with optical lasers~\cite{Ehret:2009sq}
or with microwaves in resonant cavities~\cite{Jaeckel:2007ch,Yale,Daresbury,tobar,Caspers:2009cj},
helioscopes looking for hidden photons emitted from the Sun~\cite{Gninenko:2008pz,Tokio,SHIPS},
the extragalactic X-ray diffuse background~\cite{Redondo:2008ec},
$e^+e^-$ colliders~\cite{Essig:2009nc,Batell:2009yf}, fixed-target experiments~\cite{Bjorken:2009mm,Batell:2009di},
and hadron colliders like Tevatron or the LHC~\cite{ArkaniHamed:2008qp,Baumgart:2009tn}.}
\label{probes}
\end{figure}

As sketched in Fig.~\ref{scenarios}, LARGE volume scenarios allow for a variety of different extra, hidden U(1) gauge bosons.
On LARGE cycles we can have hyperweak gauge bosons (see Fig.~\ref{scenarios}(a)), and on collapsed cycles (see Fig.~\ref{scenarios}(b)) as well as on $\overline{{\rm D}3}$-branes (see Fig.~\ref{scenarios}(c)) we can have extra U(1)s with gauge couplings comparable to those of the Standard Model gauge groups.
Masses for these hidden U(1) gauge bosons can arise either from a St\"uckelberg mechanism or from a Higgs mechanism.
The St\"uckelberg masses are typically closely linked to the type of hidden U(1) and to the
string scale.
Therefore, this situation is particularly predictive (cf. the bands labeled ``St\"uckelberg \#1" and ``St\"uckelberg \#2" in Fig.~\ref{stringexpectations}).
The hidden photon could also obtain a mass via a hidden Higgs mechanism. In these scenarios the hidden photon typically acquires
a mass comparable to the size of the soft supersymmetry breaking terms in the hidden sector. The size of the latter is strongly dependent
on the supersymmetry breaking scenario (e.g. gauge or gravity mediation) and covers a large range of possible values.
Note, however, that indeed very small masses can arise quite naturally.
In the Higgs scenario there is, of course, also a hidden Higgs particle.
This typically has a mass of the same order of magnitude as the hidden photon but this can be remedied with some fine-tuning.
Since a light hidden Higgs would in many situations behave
like a minicharged particle this puts additional constraints on this scenario: the remaining allowed region is displayed
as a shaded area labeled by ``Hidden Higgs" in Fig.~\ref{stringexpectations}.
The orange regions show allowed regions of increasing fine tuning to the right.
We note that in this particular case hidden photons with masses and mixings in the meV valley (with a very reach phenomenology, see item 1 in the Introduction) are possible.

Even smaller kinetic mixings than those depicted in Fig.~\ref{stringexpectations} can arise on $\overline{\rm D3}$ branes or
in scenarios where the visible U(1) is embedded in a GUT and the leading order contributions cancel. The typical sizes of the
kinetic mixing are shown in Fig.~\ref{Fig:chivsms_susybroken} for these cases.
Also note that in these setups the hidden gauge couplings are not necessarily hyperweak.
Moreover, despite their small values for the kinetic mixing these scenarios may nevertheless be phenomenologically interesting,
as for example in the context of decaying hidden photino dark matter (item 5 in the Introduction).

Finally, let us also note that the hidden Higgs could also remain vev-less. Then it would behave as hidden sector matter which acquires a small
electric charge under the ordinary photon due to kinetic mixing and a (positive) mass of the order of the soft mass terms ranging from $\ll {\rm eV}$ to
$\gtrsim{\rm TeV}$. The minicharges arising in the various setups discussed and the current bounds are shown in Fig.~\ref{mcps} and Fig.~\ref{mcpbounds}, respectively.
Let us also note that in hyperweak scenarios, minicharged particles arising from kinetic mixing behave in some situation as if there is
no hidden photon at all~\cite{Burrage}. This leads to some interesting phenomenological features.

In the near future a wide variety of astrophysical observations and laboratory experiments can search for hidden photons.
Comparing with our expectations from string theory (see Fig.~\ref{probes}) we find that together these experiments and observations test
a wide variety of models and an impressive range of string scales.

\section*{Acknowledgments}

MG would like to thank Karim Benakli, Joe Conlon, Nick Halmagyi, Amir Kashani-Poor, Sameer Murthy and Waldemar Schulgin for useful discussions.
JJ, JR and AR would like to thank Clare Burrage for collaboration on related subjects.
Moreover, AR would like to thank Thomas Grimm, Dieter L\"ust and Christoph Weniger.

\appendix

\section{String effective actions}\label{APPENDIX:EFFECTIVE}

Here we shall briefly review the low energy field theory of type IIB strings relevant for LARGE volume compactifications; for a more complete treatment see for example \cite{Grimm:2004uq,Jockers:2005zy}.

The starting point is the ten-dimensional bosonic action
\begin{align}
S_{\rm IIB}
=&2\pi M_s^8\int  e^{-2\Phi} \bigg(R \star 1 + 4 d \Phi \wedge \star d\Phi - \frac{1}{2} H_3 \wedge \star H_3 \bigg) \nonumber \\
&-\pi M_s^8 \int \bigg( F_1 \wedge \star F_1 + F_3 \wedge \star F_3 + \frac{1}{2} F_5 \wedge \star F_5 + C_4 \wedge H_3 \wedge F_3 \bigg),
\label{EFFECTIVE:10DIIB}
\end{align}
where in the string basis the propagating degrees of freedom are the metric variations $g_{ij}$, antisymmetric
tensor $B_{ij}$, dilaton $\Phi$ (where $g_s=e^{\langle\Phi\rangle}$) and $R-R$ $p$-forms $C_{[\mu_1,...,\mu_p]}$, and the field strengths
\beq
H_3 \equiv d B_2, \qquad F_1 \equiv d C_0, \qquad F_{p+1} \equiv d C_p - C_{p-2} \wedge H_3,
\eeq
and $F_5 = \star F_5$ must be imposed on the equations of motion.

Note that the above action (\ref{EFFECTIVE:10DIIB}) does not have a canonical Einstein-Hilbert term when the dilaton is treated as a dynamical field. One approach to remedy this is to rescale to
the ``Einstein metric'' $g_{E\mu \nu} = e^{-\Phi/2} g_{\mu \nu}$ (note that $\sqrt{g_E} R_E = \Lambda^{\frac{d-2}{2}}  \sqrt{g} R$
under $G_{E\mu \nu} = \Lambda g_{\mu \nu}$). Note that the kinetic terms of $p$-forms rescale as $e^{\frac{\Phi}{2}(4-p)}$ and the dilaton obtains a canonical kinetic term. With the new metric the volume rescales so that $V_E = e^{-\frac{3}{2} \Phi} V$.

The above is supplemented by the D-brane actions
\begin{align}
S_{{\rm D}p} =&-2\pi l_s^{-p-1}\int d^{p+1}x e^{-\Phi}{\rm tr}\sqrt{-\det\left({i^*( g+ B) + l_s^2 F/(2\pi )}\right)} \nonumber \\
&-2\pi {\rm i} l_s^{-p-1} \int_{{\rm D}p}
{\rm tr}\exp\left({i^* B+l_s^2 F/(2\pi )}\right)\wedge \sum_{m=0}^5 C_{2m} \wedge \sqrt{\frac{\hat{A}(R_T)}{\hat{A}(R_N)}}.
\label{EFFECTIVE:DBI}
\end{align}
Here $\hat{A}(R_T), \hat{A}(R_N)$ are the $\hat{A}$-roof genus associated with respectively the tangent and normal bundles on the D-brane, particularly relevant when the volume of the brane is small. For branes wrapping supersymmetric cycles (the only case we consider in this work) the first, Dirac-Born-Infeld term above becomes
\begin{equation}
S_{\rm DBI} = -\frac{2\pi}{g_s l_s^4} Z \int d^4 x {\rm tr}\sqrt{-\det\left({( g_4+ B_4) + l_s^2 F/(2\pi )}\right)}
\label{EFFECTIVE:SDBI}\end{equation}
where the integration is over the visible four dimensions (hence the subscripts denote the four-dimensional metric and antisymmetric tensor) and $Z$ is the central charge
\beq
Z = \int_{[Dp]}  \exp \bigg[ - l_s^{-2} i^*(B + i J) \bigg] \wedge \exp \bigg[ \frac{F}{2\pi} \bigg] \wedge \sqrt{\frac{\hat{A}(R_T)}{\hat{A}(R_N)}},
\eeq
where we integrate only over the compact cycle wrapped by the D$p$ brane.

The above can then be dimensionally reduced, keeping only the massless modes; we find scalars, vectors and pseudoscalars according to zero, one or two indices on the non-compact dimensions respectively. The metric variations can be decomposed in terms of four-dimensional fields as
\begin{align}
J =& i g_{i \ov{j}} d y^i \wedge d \ov{y}^{\ov{j}} = t^A (x) \omega_A,  \nonumber \\
\delta g_{ij} =& \frac{3! i}{\Omega_{ijk} \ov{\Omega}^{ijk}} \ov{z}^K (x) (\ov{\chi}_K)_{i \ov{i} \ov{j}} \Omega^{\ov{i} \ov{j}}_{\ \ j},
\end{align}
where $x$ are the coordinates on Minkowski space; $J$ is the K\"ahler form; $\omega^A$ a basis of the cohomology group $H^{1,1}$; $y^i$ some coordinates on the Calabi-Yau; $\Omega_{ijk}$ the holomorphic three-form; and $(\ov{\chi}_K)_{i \ov{i} \ov{j}}$ a basis of $H^{1,2}$. $z^K$ is a set of complex scalars; $z^K$ and $t^A$ are then the fields corresponding to variations of the Calabi-Yau metric. Note that we define all of the fields to be dimensionless. Next we examine the form potentials,
\begin{align}
B_2 =& B_2 (x) + b^A(x) \omega_A, \qquad C_2 = C_2 (x) + c^A(x) \omega_A,  \nonumber \\
C_4 =& D_2^A (x) \wedge \omega_A + V^K (x) \wedge \alpha_K - U_K (x) \wedge \beta^K + \rho_A \tilde{\omega}^A.
\end{align}
Here  $\tilde{\omega}^A$ is a basis of $H^{2,2}$; and $\alpha_K, \beta^K$ are a symplectic basis of three cycles, $K$ running from $1$ to $h^{1,2}$. We then see that the $z^K$ and $V^K$ must form the scalar and vector part of $h^{2,1}$ vector multiplets making the complex structure moduli fields, while $v^A, b^A, c^A, \rho_A$ form the four real scalars of a set of $h^{1,1}$ hypermultiplets which comprise the K\"ahler moduli.

We then introduce an orientifold projection $(-1)^{F_L} \Omega_p \sigma^*$ where $F_L$ is the left-moving worldsheet fermion number for excitations on Minkowski coordinates\footnote{So it affects only the Ramond sectors, and leaves $\Phi, g, B_2$ invariant but under which $C_0, C_2, C_4$ are odd.}; $\Omega_p$ is worldsheet parity reversal\footnote{Under which $g, \Phi, C_2$ are even and $B_2, C_0, C_4$ are odd due to the difference of GSO projections between left and right moving sectors in IIB.}; and $\sigma^*$ is the pull-back of an isometry $\sigma$ (with $\sigma^2 = 1$) of the Calabi-Yau manifold onto differential forms\footnote{Under which $\Phi, C_0, g, C_4$ are even and $B_2,C_2$ are odd.}. The action of $\sigma^*$ on the K\"ahler form is $\sigma^* J = J$ and on the holomorphic three-form $\sigma^* \Omega = - \Omega$. This generates $O3$ and $O7$ planes ( for $\sigma^* \Omega = \Omega$ we would have $O5$ and $O9$ planes). The cohomology groups $H^{p,q}$ then split into $H^{p,q}_{\pm}$ according to their eigenvalue under $\sigma^*$. We can then split $H^{1,1}$ into $H^{1,1}_+$ with basis $\omega_\alpha$ and $H^{1,1}_-$ with basis $\omega_a$ according to their eigenvalues; there is then a corresponding basis $\tilde{\omega}^\alpha, \tilde{\omega}^a$ of $H^{2,2}_+$ and $H^{2,2}_-$ respectively. Note that these are dual to the $\omega_\alpha, \omega_a$;
\begin{align}
\frac{1}{l_s^6} \int \omega_\alpha \wedge \tilde{\omega}^\beta &= \delta_\alpha^\beta ,  \nonumber \\
\frac{1}{l_s^6}\int \omega_a \wedge \tilde{\omega}^b &= \delta_a^b , \nonumber \\
\frac{1}{l_s^6} \int \omega_\alpha \wedge \tilde{\omega}^b &= 0 .
\end{align}

For the three-forms, since $\sigma$ is a holomorphic involution that commutes with the Hodge $\star$ operation, we find $h^{1,2}_{\pm} = h^{2,1}_{\pm}$. Clearly since the holomorphic three-form is unique and odd under the involution we find $H^{3,0} = H^{3,0}_-$. Of the metric variations, clearly since the K\"ahler form is even only the $H^{1,1}_+$ elements survive ($J = t^\alpha(x) \omega_\alpha$); while for the complex structure variations it is the odd elements since the holomorphic three-form is odd ($\ov{\chi}_K = \ov{\chi}_k \in H^{1,2}_-$). Then the surviving form fields are
\begin{align}
B_2 =& b^a(x) \omega_a, \qquad C_2 = c^a(x) \omega_a , \nonumber \\
C_4 =& D_2^\alpha (x) \wedge \omega_\alpha + V^\kappa (x) \wedge \alpha_\kappa + U_\kappa (x) \wedge \beta^\kappa + \rho_\alpha \tilde{\omega}^\alpha.
\end{align}
Here it is worth noting that $(-1)^F \Omega_p \sigma^* V^K, U^K = \ov{V}^K, \ov{U}^K$ and thus the invariant states $ V^\kappa ,U_\kappa$ are actually \emph{real} fields, with $(\alpha_\kappa, \beta^\kappa)$ being a real symplectic basis of $H^{2,1}_+ \oplus H^{1,2}_+$. Thus we see that half of the states have been projected out.

Note that with the orientifold projection on the forms we can define new quantities via the intersection form; on a Calabi-Yau manifold $Y$, the intersection form is defined in terms of a basis of two-forms $\{\omega_A\}, A=1..h^{1,1}$ as
\beq
K_{ABC} \equiv \frac{1}{l_s^6}\int_Y \omega_A \wedge \omega_B \wedge \omega_C.
\eeq
After the orientifold projection, the intersection numbers $K_{\alpha \beta c} = K_{abc} = K_{\alpha b c} = 0$. The volume is then given by $\mathrm{Vol}(Y) = \frac{1}{3!} \int_Y J \wedge J \wedge J \equiv \frac{l_s^6}{3!} t^\alpha t^\beta t^\gamma K_{\alpha \beta \gamma} \equiv l_s^6 \C{V}$ which involves only positive eigenforms of $\sigma^*$. The metric is also affected;
\beq
G_{AB} \equiv \frac{1}{l_s^6} \int \omega_A \wedge \star \omega_B
\eeq
and so
\begin{align}
G_{\alpha b} &= 0 , \nonumber \\
G_{\alpha \beta} &= - K_{\alpha \beta \gamma} t^\gamma + \frac{K_{\alpha \mu \nu} t^\mu t^\nu  K_{\beta \lambda \rho} t^\lambda t^\rho}{4\C{V}} , \nonumber \\
&= - K_{\alpha \beta \gamma} t^\gamma + \frac{\tau_\alpha \tau_\beta}{\C{V}} , \nonumber \\
G_{ab} &= - K_{ab\gamma} t^\gamma .
\label{EFFECTIVE:METRICS}\end{align}
Here we have defined $\tau_\alpha \equiv K_{\alpha \beta \gamma} t^\alpha t^\beta$ which corresponds to the volume (in string units) of the four-cycle dual to $\omega_\alpha$.

Now if we take the surviving fields and insert them into the effective action (\ref{EFFECTIVE:10DIIB}) we can identify the chiral superfields that are the K\"ahler coordinates on the moduli space. One finds that the K\"ahler potential  is given by\footnote{Including for completeness the $\alpha'$ corrections due to $\xi = - \frac{\chi(Y)}{2} \zeta(3)$, where $\chi(Y) = 2h^{1,1} - 2h^{1,2} $ is the Euler number of $Y$ and $\zeta$ is the Riemann zeta-function.}
\beq
K = - 2 \log [\C{V} + \frac{\xi}{2}] - \log \bigg[ -i \int \Omega \wedge \ov{\Omega} \bigg] -4 \log \bigg[\frac{1}{\sqrt{2}}( S +  \ov{S}) \bigg]
\eeq
where the complex fields, in addition to the complex structure moduli $z^k$, are
\begin{align}
S=& e^{-\Phi} - i C_0, \qquad G^a = c^a + i S b^a  , \nonumber \\
T_\alpha =& i \rho_\alpha  + \Re(S) \tau_\alpha + \frac{1}{2} \frac{1}{S+\ov{S}} K_{\alpha b c} G^b (G-\ov{G})^c.
\label{EFFECTIVE:Kaehlerdef}\end{align}
The above is given in the string frame as will appear in string effective actions from loop computations.
Noting that $\Re(\langle S\rangle ) = 1/g_s$, the appearance of $\Re(S)$ in front of the volume factor indicates that positive powers of the modulus $T_\alpha$ cannot appear perturbatively beyond tree level.

Now we note that $T_\alpha$ has a set of shift symmetries
\beq
c^a \rightarrow c^a + \theta^a, \qquad \rho^a \rightarrow \rho^a + K_{\alpha bc} b^a \theta^c
\eeq
which become gauged upon the addition of D-branes (in fact, once we add D-branes  we find that the separation between K\"ahler and complex structure moduli spaces becomes obscured, but this is not important for this work). Therefore the K\"ahler moduli, beyond tree level, may only appear as exponentials in the superpotential, and only from non-perturbative contributions. This can also be understood as the K\"ahler moduli scalars being the lowest components of linear multiplets (which may not perturbatively appear in the superpotential) whereas the complex structure moduli lie in chiral multiplets.

The complex structure moduli\footnote{Note that there are $h^{2,1}_- + 1$ of these, with the extra degree of freedom corresponding to shifts of the holomorphic three-form $\Omega \rightarrow \Omega e^{-h}$; this is removed by choosing a basis $z^k = (1,z^k)$.} $z^k$ do not have a shift symmetry and do not carry factors of the string coupling, and therefore the superpotential and gauge kinetic functions may be arbitrary functions of these. This is in contrast to the case in IIA strings, where both complex and K\"ahler moduli have Peccei-Quinn shift symmetries.

\section{$\cancel{\rm SUSY}$ kinetic mixing terms in field theory}\label{susyfield}

Here we wish to calculate the operators (\ref{SPECIAL:secondorderF},\ref{SPECIAL:secondorderD}) for a generic renormalisable model having messenger superfields $\Phi_i = \phi_i + \sqrt{2} \theta^\alpha \psi_{i\, \alpha} + (\theta \theta) F_i + ...$ coupling to a pseudomodulus $\Xi$ having a vev $\langle \Xi \rangle = F_\Xi (\theta \theta)$. The relevant superpotential coupling is
\beq
W = W_{\Xi i j} \Xi \Phi_i \Phi_j.
\eeq

We shall need the formula
\beq
\int \frac{d^4 q}{(2\pi)^4} \prod_{i=1}^4 \frac{1}{q^2 - m_i^2} = - \frac{1}{16\pi^2} \sum_{i \ne j \ne k \ne l}^4 \frac{1}{6} \frac{m_i^2 \log m_i^2}{(m_i^2 - m_j^2)(m_i^2-m_k^2)(m_i^2 - m_l^2)}
\eeq
and the simplifications
\begin{align}
\int \frac{d^4 q}{(2\pi)^4} \frac{1}{(q^2 - m^2)^4} &= \frac{1}{16\pi^2}\frac{1}{6m^4}  , \\
\int \frac{d^4 q}{(2\pi)^4} \frac{1}{(q^2 - m^2_i)} \frac{1}{(q^2 - m^2_j)^3} &\equiv f_{ij}^{(1,3)} = \frac{1}{16\pi^2} \bigg[\frac{1}{2m_j^2 (m_j^2 - m_i^2)} + \frac{m_i^2}{(m_j^2 - m_i^2)^3} \log \frac{m_i^2}{m_j^2}\bigg] , \nonumber \\
\int \frac{d^4 q}{(2\pi)^4} \frac{1}{(q^2 - m^2_i)^2} \frac{1}{(q^2 - m^2_j)^2} &\equiv f_{ij}^{(2,2)} = \frac{1}{16\pi^2} \bigg[-\frac{2}{ (m_j^2 - m_i^2)^2} + \frac{m_i^2 +m_j^2}{(m_j^2 - m_i^2)^3} \log \frac{m_i^2}{m_j^2}\bigg] . \nonumber \\
\end{align}
To compute $\cancel{\rm SUSY}$ kinetic mixing terms, the most efficient method is to compute the ``scattering'' of the auxiliary fields in a manner similar to ``generalised kinetic mixing'' of \cite{Benakli:2008pg},
and so to compute terms of the form \eqref{SPECIAL:firstorder},\eqref{SPECIAL:secondorderF},\eqref{SPECIAL:secondorderD} we compute the
scattering of two $D$ fields with one and two $F$ fields, and two $D$ fields respectively. Let us consider the couplings of these to
messenger fields $\Phi_i$ with masses $m_i$ (we can always choose a basis such that the mass matrix is diagonal); the $D$ terms couple
via their gauge current $ \sqrt{2} D q_{i} \phi_i^\dagger \phi_i$,\footnote{In principle after diagonalising the (supersymmetric) mass matrix we may obtain non-diagonal gauge currents. However, this only occurs when we allow vevs of fields that would spontaneously break the gauge symmetry - since we are only interested in the (gauge invariant) $F$- and $D$-term contributions we can safely ignore these.}
while the $F$-terms couple via the superpotential $F_\Xi W_{\Xi ij} \phi_i \phi_j$.

We then find effective terms in the Lagrangian:
\begin{align}
-\Delta \C{L} \supset & \frac{1}{16\pi^2} 16 f_{ij}^{(1,3)} W_{\Xi i j} W^{\dagger}_{\ov{\Xi} i j} q_{j} q_{j}^{\prime}   D D' |F_{\Xi}|^2 \nonumber \\
&+ \frac{1}{16\pi^2} 8 f_{ij}^{(2,2)} W_{\Xi i j} W^{\dagger}_{\ov{\Xi} i j} q_{i} q_{j}^{\prime}  D D' |F_{\Xi}|^2 \nonumber \\
& \sum_i \bigg(\frac{1}{16\pi^2} \frac{1}{m_i^4} (q_i)^2 D^2 (q_i')^2 (D')^2 + \frac{1}{16\pi^2} \frac{2}{3m_i^4} q_i D (q_i')^3 (D')^3 + \frac{1}{16\pi^2} \frac{2}{3m_i^4} (q_i)^3 D^3 q_i' D'\bigg) .
\end{align}
However, to find the corresponding terms in the K\"ahler potential for the $F$-terms we must be careful, since there may be terms of the form
\beq
\int d^4 \theta W^\alpha (D_\alpha \Xi) \ov{W}_{\dot{\alpha}}^\prime \ov{D}^{\dot{\alpha}} (\ov{\Xi}) \supset 4 D D' |F_\Xi|^2 ,
\eeq
which contains no kinetic mixing term. Therefore we must compare with the scattering of a photon and a hidden photon with two auxiliary $F$ fields and examine the coefficient of $p_\mu p_\nu^\prime$. Since the auxiliary fields do not couple to the fermions, the photon vertex coupling to scalars is identical to the coupling of the $D$ field to scalars up to a factor of the external momentum; therefore we do not generate terms of the above form, and we conclude that the corrections to the K\"ahler potential are
\begin{align}
-\Delta K \supset & -\sum_{ij}\frac{1}{16\pi^2} W_{\Xi i j} W^{\dagger}_{\ov{\Xi} i j} \bigg[2 f_{ij}^{(1,3)}  q_{j} q_{j}^{\prime} + f_{ij}^{(2,2)}  q_{i} q_{j}^{\prime} \bigg]\bigg( W^{\alpha} W_{\alpha}^\prime \ov{\Xi} D^2 (\Xi)  + \ov{W}_{\dot{\alpha}} \ov{W}^{\prime \dot{\alpha}} \Xi \ov{D}^2 (\ov{\Xi}) \bigg)\nonumber \\
&+ \sum_i \frac{1}{16\pi^2} \frac{1}{m^4_i} q_{i} q_{i}^{\prime} q_{i} q_{i}^{\prime}  W^\alpha W_{\alpha}^\prime  \ov{W}_{\dot{\alpha}} \ov{W}^{\prime \dot{\alpha}} \nonumber \\
&+ \sum_i\frac{1}{16\pi^2} \frac{1}{3m^4_i} q_{i} q_{i} q_{i} q_{i}^{\prime}  (W^\alpha W_{\alpha} \ov{W}_{\dot{\alpha}} \ov{W}^{\prime \dot{\alpha}} + W^\alpha W_{\alpha}^\prime \ov{W}_{\dot{\alpha}} \ov{W}^{ \dot{\alpha}}) \nonumber \\
&+\sum_i \frac{1}{16\pi^2} \frac{1}{3m^4_i} q_{i} q_{i}^\prime q_{i}^\prime q_{i}^{\prime}( W^\alpha W_{\alpha}^\prime \ov{W}_{\dot{\alpha}}^\prime \ov{W}^{\prime \dot{\alpha}} + W^{\alpha \prime} W_{\alpha}^\prime \ov{W}_{\dot{\alpha}}^\prime \ov{W}^{ \dot{\alpha}}).
\end{align}

\section{$\cancel{\rm SUSY}$ kinetic mixing terms in toroidal string models}
\label{APPENDIX:TOROIDAL}

To calculate the terms in the effective action \eqref{SPECIAL:firstorder},\eqref{SPECIAL:secondorderF},\eqref{SPECIAL:secondorderD} for a toroidal model, we can compute scattering amplitudes with the auxiliary fields. We shall consider an orientifold of a factorisable six-torus $\mathbb{T}^6 = \mathbb{T}^2_1 \times \mathbb{T}^2_2 \times \mathbb{T}^3_3$. The vertex operator for the $F$-term for a K\"ahler modulus corresponing to the $j^{\mathrm{th}}$ torus is
\beq
V_{F}^{(0,0)} =  \frac{e^{ik \cdot X}}{T^j - \ov{T}^j}  \frac{1}{\sqrt{2\alpha'}}\bigg[\partial X^j \prod_{\kappa \ne j}^3 \tilde{\Psi}^\kappa  + \ov{\partial} X^j \prod_{\kappa \ne j}^3 \ov{\Psi}^\kappa \bigg] ,
\eeq
where $T^j$ is the magnitude of the K\"ahler modulus, and, as mentioned in the text the operator contains no Chan-Paton factors and so will amplitudes involving it will cancel for traceless hypercharge and $N=4$ sectors. The operator for the $D$-term is
\beq
V_D^0 = \lambda e^{ik \cdot X} \sum_{\kappa = 1}^3 \partial H_k (z) = \lambda  e^{ik \cdot X} \sum_{\kappa = 1}^3 \lim_{w \rightarrow z} -i \partial_w \bigg[(w-z)e^{i H_\kappa (w)} e^{-i H_\kappa (z)}\bigg] ,
\eeq
where $\lambda$ is the Chan-Paton factor.

However, we can calculate the coefficient of the operator $W^a W^b W^c W^d$ where $a,b,c,d \in \{h,\mathrm{vis}\}$ by considering
\begin{align}
\int d^4 \theta W^a W^b \ov{W}^c \ov{W}^d \supset& D^a D^b D^c D^d - D^a D^b \bigg( \frac{i}{4} F^c_{\kappa \rho} \tilde{F}^{c\ \kappa \rho} + \frac{1}{2}F^c_{\kappa \rho} F^{d\ \kappa \rho} \bigg) \nonumber \\
&- D^c D^d \bigg( \frac{i}{4} F^a_{\mu \nu} \tilde{F}^{b\ \mu \nu} + \frac{1}{2} F^a_{\mu \nu} F^{b\ \mu \nu} \bigg) \nonumber \\
&+\bigg( \frac{i}{4} F^a_{\mu \nu} \tilde{F}^{b\ \mu \nu} + \frac{1}{2} F^a_{\mu \nu} F^{b\ \mu \nu} \bigg)\bigg( \frac{i}{4} F^c_{\kappa \rho} \tilde{F}^{c\ \kappa \rho} + \frac{1}{2}F^c_{\kappa \rho} F^{d\ \kappa \rho} \bigg) ,
\end{align}
and so by extracting the coefficient of $\frac{1}{4} F^a_{\mu \nu} F^{b\ \mu \nu}F^c_{\kappa \rho} F^{d\ \kappa \rho}$ in a scattering amplitude we can determine the coefficient in the effective action. The result of \cite{Bianchi:2006nf} gives
\begin{align}
\lim_{p_a,p_b,p_c,p_d \rightarrow 0} A (a,b,c,d) &\supset \frac{(2\pi)^4 }{4N} \left(\frac{-1}{4}  F^a_{\mu \nu} F^{b\ \mu \nu}F^c_{\kappa \rho} F^{d\ \kappa \rho}\right)  \int_0^\infty dt \frac{8(2\alpha')^4 t}{(8\pi^2\alpha')^2} \Lambda (t) ,
\end{align}
where $p_i$ are the momenta, $A$ is the amplitude, $N$ is the order of the orbifold (the prefactor $4$ is from the GSO and orientifold projections) and $\Lambda$ is the momentum/winding sum. The above assumes that the branes are separated so that the amplitude is regulated in the infra-red of the field theory; this is the case we are interested in. Consider for simplicity a rectangular six-torus of equal radii $R$, and brane separations along each coordinate of $y_i$. Then the winding sum is
\beq
\Lambda (t) = \prod_{i=1}^6 \sum_{n_i} e^{-\frac{t}{2\pi\alpha'}(2\pi R n_i + y_i)^2} = \frac{1}{(2t)^3} \frac{1}{\C{V}} \prod_{i=1}^6 \sum_{m_i} \exp[ -\frac{\pi \alpha' m_i^2}{2tR^2} - 2\pi i m_i \tilde{y}_i ] ,
\eeq
where we defined $\tilde{y}_i \equiv y_i /2\pi R$, $n_i, m_i$ are integers and of course $\C{V}= \frac{(2\pi R)^6}{(4\pi^2\alpha')^3}$. This is then very similar to brane-antibrane kinetic mixing.
Performing the integral in the closed string channel and excluding the zero mode corresponding to a massless closed string exchange gives
\beq
c_D = \frac{1}{2\pi^4}  \frac{2}{\pi N\C{V}^{2/3}}  \sum_{m_i \ge 0} \frac{\cos (2\pi \sum_{j=1}^6 m_j \tilde{y}_j)}{\sum_{j=1}^6 m_j^2}.
\eeq
However, the appropriate estimate in \cite{Abel:2003ue} involves integrating in the open string channel and neglecting the winding modes; this yields
\beq
c_D \approx \frac{1}{2\pi^4} \frac{1}{2\pi} \frac{1}{N} \frac{1}{\C{V}^{2/3}} \frac{1}{(\sum_i\tilde{y}_i^2)^2} .
\eeq
However, this is not periodic in $\tilde{y}_i$; therefore we shall simply take as our estimate in general to be
\beq
c_D \approx \frac{1}{4\pi^5} \frac{1}{ \C{V}^{2/3}}.
\eeq

\end{document}